\documentclass{aa} 

\usepackage{graphicx}

\usepackage{longtable,lscape} 

\usepackage{lscape} 

\usepackage{txfonts}
\def\msun{\mbox{{\rm M}$_\odot$}}

\begin{document}

\title{PNe as observational constraints in chemical evolution models for NGC 6822.} 
\author{
Liliana Hern\'andez-Mart\'inez\inst{1,2}, 
Leticia Carigi\inst{1}, Miriam Pe\~na\inst{1}, and Manuel Peimbert\inst{1}} 

\offprints{L. Hern\'andez-Mart\'inez} 

\institute{
1 Instituto de Astronom\'ia, Universidad Nacional Aut\'onoma de M\'exico,
Apdo. Postal 70264, M\'ex. D. F., 04510 M\'exico.\\ 
2 Instituto Nacional de Astrof\'isica, Optica y Electr\'onica, Luis
Enrique Erro No. 1, Puebla, M\'exico.\\ 
\email{lilihe@inaoep.mx;carigi,miriam,peimbert@astroscu.unam.mx} } 
\date{Received xxxxxxxxxxx; accepted xxxxxxxxxxxx} 

\titlerunning{Chemical enrichment in NGC\,6822}

\authorrunning{Hern\'andez-Mart\'inez et al.} 


\abstract 
{} 
{
Chemical evolution models  are useful for understanding the 
formation and evolution of stars and galaxies. 
Model predictions will be more robust as more observational constraints  are used.  
We present chemical evolution
models for the dwarf irregular galaxy NGC~6822  using chemical
abundances of  old and young Planetary Nebulae (PNe) and \ion{H}{ii} regions
as  observational  constraints. Two sets of chemical abundances,
one derived from collisionally excited lines (CELs) and one, from
recombination lines (RLs), are used.  We  try
to use our  models as a tool to discriminate between both procedures for abundance determinations.
}
{
In our chemical evolution code, the chemical contribution of low and
intermediate mass stars is time delayed, while  for the massive stars
the chemical contribution follows the instantaneous recycling approximation. 
Our models have two main free parameters: 
 the mass-loss rate of a well-mixed outflow and 
the upper mass limit, $M_{up}$, 
of the initial mass function (IMF). 
To reproduce the gaseous mass and the present-day O/H value 
 we need to vary the outflow rate and the $M_{up}$ value.
}
{ 
We calculate two models with different $M_{up}$
values that reproduce adequately the constraints. 
The abundances of old PNe are in  agreement with our models and support  the star formation history derived independently from photometric data.  
In addition, by assuming a fraction of binaries producing SNIa of 1\%,  the models 
fit the Fe/H abundance ratio as derived from A supergiants.
The first model (M4C), that assumes $M_{up}=40$ \msun,
fits, within errors smaller than  2 $\sigma$, the  O/H, Ne/H, S/H, Ar/H and Cl/H abundances obtained from CELs, 
 for old and young PNe and \ion{H}{ii} regions.
The second model (M1R), that adopts $M_{up}=80$ \msun, reproduces, within 2 $\sigma$ errors, 
the O/H, C/H, Ne/H and S/H abundances adopted from RLs.
Model M1R does not provide a 
good fit to the Cl/H and Ar/H ratios,
because the SN yields of those elements for $m >40$ \msun \ are not adequate and need to be improved (two sets of yields were tried). 
From these results we are not able to conclude which set of abundances 
(the one from CELs or the one from RLS) represents better the real abundances in the ISM.
We discuss the predicted $\Delta Y/\Delta O$ values, finding that the value from 
model M1R  agrees better with data for other galaxies from the literature than
the value from model M4C.
} 
{}


\maketitle 


\section{Introduction}

Planetary Nebulae (PNe) constitute one of the most valuable chemical
tracers of the past abundances in the interstellar medium (ISM). Their
chemical compositions allow us to determine  the abundances of
some chemical elements present in the ISM when their progenitor stars
were born. PNe are produced by stars with initial masses from
$\sim$  1 \msun \ to $\sim$  8 \msun \ and also with a large
age spread (from 0.1  to 9 Gyr, Allen et al. 1998). Therefore PN
characteristics  are important as observational constraints in chemical
evolution models, allowing us  to improve the inferred chemical history
(Hern\'andez-Mart\'inez et al. 2009, hereafter HPCG09; Richer \& McCall
2007; Buzzoni et al. 2006; Maciel et al. 2006).
Despite the fact that PNe show some bright  emission lines, deep
observations are needed to  determine their physical conditions and
accurate chemical abundances, which are based on much fainter lines.

In addition, gaseous nebulae are an important key in the chemical abundances determination
of noble gases (e.g., Ne and Ar) and other elements like Cl
 and, therefore, in the test of stellar yields of these elements.
The determination of this type of elements in stars is not so reliable
and, in previous papers (e. g., Timmes et al. 1995; Romano et al. 2010; Kobayashi et al. 2011), 
the authors were not able to test  Ne, Cl, and Ar yields, due to the lack of stellar abundances.

The chemical evolution equations (e.g., Tinsley 1974) take into account
many physical parameters: galactic infalls, galactic outflows, the initial
mass function, the star formation rate, and a set of stellar yields for
different masses. Therefore these equations are complex and have to be
solved using numerical methods. However they can be simplified assuming 
the {\em instantaneous recycling
approximation} (IRA, Talbot \& Arnett  1971). For this approximation
the lifetimes of all stars more massive than 1 \msun \ are negligible compared with the age of the
galaxies. This approximation allows us to solve the chemical evolution
equations analytically. Despite its simplicity, IRA is a good first approximation
for elements produced mainly by massive stars (MS), but not for the
elements produced partially by low and intermediate mass stars (LIMS).

There are intermediate methods to calculate chemical evolution models,
which consist in  analytical approximations that consider the delays in
chemical enrichment produced by LIMS.  Several authors have presented
their own analytical approximations (e.g., Serrano \& Peimbert 1983;
Pagel 1989; Franco \& Carigi 2008).  They propose some time-delay
prescription for the chemical enrichment produced
by LIMS. These time-delay terms make the LIMS to bring out to
the ISM the processed nuclear material at a single time after their
formation, while the contribution due to MS is instantaneous, like in  the
IRA approximation.

In this paper we calculate chemical evolution models for the dwarf
irregular galaxy NGC\,6822 following the method used by  Franco \& Carigi
(2008).  This method was modified by Hern\'andez-Mart{\'\i}nez (2010)
to include numerically infalls, outflows, and star formation rates. 
 Also, in this new code we have increased the number of chemical elements 
considered, from 5 to 27.

NGC 6822, a galaxy of the Local Group, is located at 460 kpc from
the Milky Way (Gieren et al. 2006). It presents a recent increase in the
star formation rate   as shown by its bright \ion{H}{ii} regions. These
features make it easy to determine the present-day chemical abundances
of the ISM. Thus, it is suitable for chemical evolution modeling.

Carigi et al. (2006, hereafter CCP06) performed chemical and photometric
evolution models for NGC\,6822.  Based on a cosmological approach they
obtained the gas infall rate adequate to form the galaxy and, based on the 
photometric properties, they derived a robust star formation history. 
Their chemical evolution models
were built to reproduce the present component of the ISM, as given by
the chemical abundances of the \ion{H}{ii} region HV, determined from
recombination lines (RLs).

HPCG09 determined abundances from collisionally excited lines (CELs)
for 11 PNe and one \ion{H}{ii} region.  From these data, they confirmed
the chemical homogeneity of the present component in the ISM  and found
the presence of two populations of PNe. Based on these results  they
built a preliminary chemical evolution model to reproduce the chemical
behavior  of NGC\,6822. One of the aims of this work is to compute
detailed chemical evolution models by using the abundances of the ISM past
component (represented by PNe) as an additional restriction to the one
posed by \ion{H}{ii} regions to the ISM current component.

Moreover, as it is known that abundances obtained from RLs are larger
that those from CELs, by factors of about 2, a problem  known
as ``the abundance discrepancy'' (see \S5),  we also  compute models to reproduce
the chemical abundances as derived from RLs. Then, by means of these
models we  try to discriminate between the values of abundances obtained
from both methods.

Also, chemical evolution models can be used to constrain the stellar yields,
when the model results are compared with accurate observations.
In this work, we try to fit the observed Cl and Ar abundances, and derived SN yields
for massive stars.

Furthermore, we also compute the He to O abundance enrichment ratio,
$\Delta${\it Y}/$\Delta${\it O} (in this expression {\it Y} and {\it O}
are the chemical abundances of He and O, both by mass), which is an important
parameter in chemical evolution discussions.

This paper is organized as follows: In \S2 we describe the chemical
evolution equations used by assuming a time-delay approximation for LIMS.
The assumptions for the models are presented
in \S3. In \S4  and \S5 we present the results from our models, with
observational restrictions obtained from CELs and RLs, respectively. In
\S6 we discuss the evolution of $\Delta${\it Y}  vs. $\Delta${\it O}. 
Finally, in \S7 we discuss our results and present our conclusions.


\section{Chemical evolution equations with a delayed contribution
from LIMS}

Franco \& Carigi (2008) proposed an analytical solution
for the chemical evolution equations, by considering that the enrichment
produced by LIMS is delayed by
the lifetime, $\tau$, of a representative mass, while the MS enrich 
instantaneously the ISM. This solution was built only for a star formation
rate proportional to the gas mass in a closed box model. They obtained
results very similar to those found from numerical models that consider
the lifetime of each star. Franco \& Carigi studied the evolution of five elements:
H, He, C, N, O, and the metallicity {\it Z}.

Following the prescription given by Franco \& Carigi (2008),  we developed a numerical code
which solves the  differential equations of the evolution of the gas mass
${M_{gas}} (t)$ (eq. 1), and the evolution of the chemical abundance by mass 
in the ISM for the element {\it i}, $X_i(t)$ (eq. 2), respectively (Hern\'andez-Mart{\'\i}nez 2010).

This numerical code can solve 27 chemical species (H, He, C, N, O, F,
Ne, Na, Mg, Al, Si, P, S, Cl, Ar, K, Ca, Sc, Ti, V, Cr, Mn, Fe, Co,
Ni, Cu, Zn) for which we have a good collection of {\it Z} dependent
integrated yields for LIMS (y$_{lims}$), MS (y$_{ms}$) and Type Ia
supernovae, SNIa (y$_{snia}$), for 6 initial stellar metallicities
($Z_i$ = 1.0$\times$10$^{-8}$, 1.0$\times$10$^{-5}$, 1.0$\times$10$^{-4}$,
4.0$\times$10$^{-3}$, 8.0$\times$10$^{-3}$, and 2.0$\times$10$^{-2}$).
These yields are described in detail  in section 3.

In the delayed approximation the evolution of the gaseous mass is
given by:

\begin{equation}  
\frac{dM_{gas}(t)}{dt}=-(1-{\rm R}_{ms})\psi(t)+{\rm R}_{lims}\psi(t-\tau_m)+f(t)-w(t),
\end{equation}

\noindent here, R$_{ms}$ and R$_{lims}$ are the masses returned to the
ISM by massive and low-intermediate mass stars, respectively.  $\tau_m$
is the time delay of LIMS for their total gas ejection to the ISM,
and $\psi$(t), $f$(t) and $w$(t) are the star formation, accretion and
outflow rates as function of time, respectively.

In addition, the evolution of the gaseous mass in element $i$ is:

\begin{equation} \label{dxi} 
\frac{d(X_i(t)
M_{gas}(t))}{dt}=-X_i(t)\psi(t)+E_i(t)+X_i^f(t) f(t)-X_i^w(t) w(t),
\end{equation}

\noindent  where, 
$X^f_i$(t) and $X^w_i$(t) are the abundances (by mass) of element $i$
 in the gas accreted and  in the gas expelled from the
galaxy,  respectively.
$E_i(t)$ is the gas rate,  in the element {\it i}, returned  to the
ISM by MS, LIMS, and SNIa that die at a time $t$,  and it is calculated by,

\begin{displaymath} \label{ei1} 
E_i=[{\rm R}_{ms} X_i(t) + {\rm y}_{i,ms}]\psi(t)+[{\rm R}_{lims} X_i(t-\tau_i) +
{\rm y}_{i,lims}]\psi(t-\tau_i) + 
\end{displaymath} \begin{displaymath}
\label{ei2} 
\;\;\;\;\;\;\;\;\;\;{\rm y}_{i,snia}\psi(t-\tau_{snia}),
\end{displaymath}

\noindent here, $\tau_i$ is the time delay when the group of LIMS  enrich
the ISM with the element $i$ (see Franco \& Carigi 2008) and
$\tau_{snia}$ is the time delay of SNIa to pollute the gas mass of the galaxy.

In Appendix A we present, for the six initial stellar
metallicities and two different upper mass limits of the initial mass function,
$\tau_m$, $\tau_i$,  R$_{ms}$, R$_{lims}$, y$_{ms}$, y$_{lims}$, and y$_{snia}$ 
for the elements considered in this work.


\section{Assumptions of the Chemical Evolution Models}

In the following we describe in detail the assumptions included in  our models.

\begin{figure}[ht] \begin{center}
\includegraphics[width=\columnwidth]{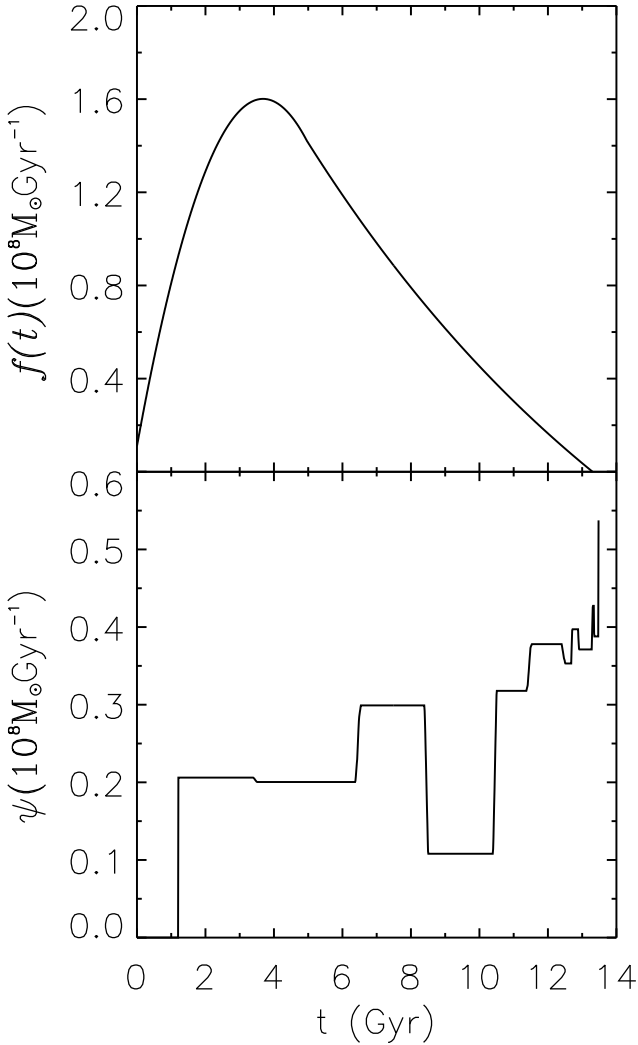} \caption{
Accretion (upper panel, see eq. 3) and star formation  (bottom panel, CCP06) histories
adopted in the models. 13.5 Gyr corresponds to the present time
 } \label{varfis} \end{center}
\end{figure}

1) We adopt  the  mass accretion rate  proposed by CCP06, which was
tailored to NGC 6822 from cosmological models of galaxy formation. 
Therefore, the accretion history is 
a  better approximation for this galaxy than simple parametric equations. 
CCP06 described two families of cosmological models for the increase of baryonic mass: 
S (small-infall) and L (large-infall). We have chosen the
L-models because they adjust better the observational
constrains presented by CCP06 (the S-models predict higher enrichment
in O/H and C/H than the L-models). In addition, we have parametrized
the accretion history of the baryonic mass as a function of time,
derived by CCP06. Such a parametrization,  which is used for our chemical
evolution models, is presented in Fig.~\ref{varfis}  (upper panel) and it is given by
the expression: 

\begin{equation} 
\label{facc} 
f(t) = 131.4(e^{-0.31t} - e^{-0.32t}) \;\;\;\;\;\;(10^8 \,\msun \,${\rm Gyr}$^{-1}). 
\end{equation}

The accreted material is assumed to be primordial, that is, with a
chemical composition given by: $X$=0.752, $Y$=0.248, and $Z=0.000$ as derived from
WMAP observations by Dunkley et al. (2009) and for metal poor irregular
galaxies by Peimbert et al. (2007a).

\smallskip

2) We  consider the history of the star formation rate  (SFR) derived for  NGC 6822
by CCP06 based on observed color-magnitude  diagrams and photometric
evolution models (Fig.~\ref{varfis}, bottom panel). 
 Interestingly it is observed that
 the accretion mass history (shown in the upper panel of Fig.~\ref{varfis}) and  the SFR do not follow the same trend 
and seem to be in contradiction in the latter 3 or 4 Gyr, when the accretion diminishes while the SFR increases  abruptly. 
It should be considered however that the SFR should not follow the gas accretion closely as this accretion is only one of the mechanisms for increasing the SFR. 
On large scales, the formation of stars is controlled by an interplay between self-gravity and supersonic turbulence, operating on the gas already accreted. 
It is not clear how star formation bursts are triggered in dwarf galaxies but, for NGC\,6822, Gouliermis et al. (2010) suggest that turbulence on kpc scales is the major agent regulating star formation. 
Several mechanisms have been proposed as able to produce   turbulence in large scales.  
In particular, for NGC\,6822,  a possible mechanism, that could have triggered at least the increase in the SFR during the last 100-200 Myr,  
has been found in the presence of  an apparently independent stellar structure called the ``northwestern'' companion, 
which is considered to be currently interacting with the main body of NGC 6822 and causing the tidal arms in the southeast of the galaxy (de Block  \& Walter 2000, 2003).
 If such an interaction occurred in the past (which is possible if the northwestern companion is a satellite nearby NGC 6822) 
it could have been the mechanism for triggering different starbursts in NGC\,6822.

\smallskip

3) We use the initial mass function (IMF) of Kroupa et al. (1993), 
in the $7.5 - M_{up}$ \msun \ mass range for $Z < 10^{-5}$ and
in the $0.1 - M_{up}$ \msun \ mass range for $Z > 10^{-5}$.
$M_{up}$ is a free parameter adjusted to reproduce the observed O/H abundance ratio.
\smallskip

4) We assume different stellar yields, dependent on initial stellar mass ($m$) and on
initial stellar metallicity $Z_i$,
provided by different authors. However they do not cover all the mass range. In the case of
LIMS, we have a complete library from 1\,M$_\odot$ to 6\,M$_\odot$ and
for MS, the mass range covered goes from 9\,M$_\odot$ to 80 M$_\odot$.
In order to fill the mass gap (from 6\,M$_\odot$ to 9\,M$\odot$), we
proceeded as follows: 
for the $6 < m/M_\odot \leq 7.5 $ mass range we assigned the stellar yield values 
given by mass fraction ($p_i$)
of the  6\,M$_\odot$ star.  In the same way, for the $7.5 <
m/M_\odot < 9$ mass range we assigned the $p_i$ values of the 9\,M$_\odot$ star.

Specifically,

\hskip 1cm  4.1) For LIMS we adopt the stellar yields  by Karakas \&
Lattanzio (2007), and the integrated yields used by our model are
given by:

\begin{equation}  
{\rm y}_{i,lims}=\int^{7.5 M_\odot}_{1 M_\odot}
mp_i(m) IMF(m) dm.
\end{equation}

\noindent
The results for ten elements,  five initial stellar metallicities, and two different $M_{up}$ values
are presented in Tables 4 and 5.

\hskip 1cm 4.2) For MS,  we consider the stellar yields obtained from
models for the pre-SN stage and the SN stage. The pre-SN yields are
taken from the work of the Geneva group (Maeder 1992; Meynet \& Maeder 2002;
Hirschi et al. 2005; Hirschi 2007). The SN yields are taken from Woosley \&
Weaver (1995) adopting their models B for the 12 - 30\,M$_\odot$
range and their models C, for the 35 - 40\,M$_\odot$ range. We combine
the Geneva group  yields with the Woosley \& Weaver  yields using the
prescription proposed by Carigi \& Hernandez (2008) where they connect
the mass of the carbon-oxygen cores (M$_{CO}$) from the Geneva group to
M$_{CO}$ from Woosley \& Weaver, using the prescription given by Portinari et al. (1998).  
Under these assumptions the He, C, N, and O yields are
equal to the pre-SN ones, but for heavier elements the yields
are similar to the SN ones. For $m > 40$ \msun \ the
adopted yields for the heavier elements are similar to those given
for $m = 40$ \msun.

The chemical contribution of MS is in IRA, and the integrated yields
are given by:

\begin{equation} 
{\rm y}_{i,ms}=\int^{M_{up}}_{7.5 M_\odot}
mp_i(m)IMF(m)dm. 
\end{equation}

\noindent
The results for ten elements,  six initial stellar metallicities, and two different $M_{up}$ values
are presented in Tables 4 and 5.

\hskip 1cm 4.3) For SNIa, we consider
the stellar yields, P$_i$(snia), given in \msun \ by Nomoto et al. (1997) which are
almost independent of the initial stellar metallicity.  We assume that
a fraction A$_{bin}$ of the stars with masses between 3 and 15\,M$_\odot$
corresponds to binary systems and that every one of these systems becomes
a SNIa. A$_{bin}$ is a free parameter of the model which is found by
fitting the observed Fe/H ratio. It is considered that all SNIa  of each stellar
generation enrich the interstellar medium at $\tau_{snia}$=1~Gyr after
the formation of SNIa progenitors (see Fig. 1 of Mannucci 2008). The
integrated yields are given by:

\begin{equation}  
{\rm y}_{i,snia}={\rm A}_{bin}P_i(snia)\int^{15
M_\odot}_{3 M_\odot} IMF(M_B)dM_B, \end{equation}

\noindent where $M_B$ is the mass of the binary system.
${\rm y}_{i,snia}$ values for ten $i$-elements
and two different $M_{up}$  values
are presented in Tables 4 and 5.
\smallskip

5)  The evolution of all the models is followed  from an initial
time, $t_0$=0, to a final time, $t_f$=13.5Gyr, with a step of
$\Delta t$=0.01Gyr. The initial conditions in all the models are
$M_{gas}$(t$_0$)=0 and $M_{star}$(t$_0$)=0.
\smallskip

6) For the outflow rates, we use  two types of prescriptions:

\hskip 1cm 6.1) Well-mixed winds. We assume that the ejecta of MS, LIMS,
and SNIa  are well-mixed with the ISM of the galaxy before part of the ISM is expelled
to the intergalactic medium (IGM). In this work, we consider that well
mixed winds depend on the SFR and consequently  the outflow rate is given by,

\begin{equation} \label{windwell} w(t)=\nu \psi(t), \end{equation}

\noindent where $\nu$ is a free parameter in our models, adjusted to
reproduce the observed $M_{gas}$ of the galaxy.

\hskip 1cm 6.2) Selective winds. We assume that during the time interval when a selective wind 
lasts, the ejecta produced by MS during that interval is expelled from the galaxy 
to the IGM without mixing with the ISM, therefore these ejecta do not contribute to the chemical enrichment
of NGC\,6822. The total mass lost due to the selective winds is small
relative to the gaseous mass of the galaxy.


\section{Models for abundances derived from collisionally excited lines}


\subsection{Observational constraints}

NGC 6822 has been well studied  along the years by different authors, therefore good observational
constraints can be found in the literature. For instance, CCP0 derived its total dynamic, total baryonic, 
and gaseous masses.
They amount to $M_T=($2.0$\pm$0.5)$\times$10$^{10}$~M$_\odot$,
$M_{bar}=($4.3.$\pm$0.2)$\times$10$^{8}$~M$_\odot$ and
$M_{gas}$=(1.98$\pm$0.2)$\times$10$^8$~M$_\odot$, respectively. 

The chemical abundances of several \ion{H}{ii} regions, determined by  diverse
authors  from collisionally excited lines  (not corrected for dust depletion) were compiled by HPCG09, 
with the exception of Cl/H that comes from Peimbert et al. (2005).
 The average O/H abundance of those
\ion{H}{ii} regions is 12+log(O/H)$=8.08\pm0.05$, and HPCG09  found
no evidence of chemical inhomogeneities in this galaxy for the
\ion{H}{ii} regions located  in a central  area of about 3 kpc in
diameter. In this work we will include a correction due to depletion
in dust grains, by adding  0.08 dex to the O gaseous
abundances of \ion{H}{ii} regions, correction suggested by Peimbert et
al. (2005) as adequate for  NGC\,6822. From this we obtain 12+log(O/H)$=8.16 \pm0.05$.
On the other hand,  from  A-type supergiant 
stars, Venn et al. (2001) derived the chemical abundance of iron to be 12 + log(Fe/H)=$7.01\pm$0.20.  
These data will be used as observational constraints for the present 
component of the ISM.

In addition, HPCG09 obtained the chemical  abundances of 11
PNe. They found that some of them are relatively young and their
average O/H ratio (12+log(O/H)$=8.11\pm0.10$) is  similar to the one of  \ion{H}{ii} regions. 
We tag these PNe as 'young PNe'. Besides, there is a sample  of PNe with
much lower O, showing an average of 12 + log(O/H)\,=\,7.72$\pm$0.10.
We tag these PNe as 'old PNe'. The  chemical abundances  of both groups will be used
in the chemical evolution models, as additional constraints corresponding
to the past and recent-past component of the ISM.  No correction for dust is considered
for PNe.

At the bottom of Table ~\ref{modcel} we present the  O/H, N/O, Ne/O,
S/O, Cl/O, and Ar/O abundance ratio  averages for the best determined \ion{H}{ii} regions
(the correction for dust is included), young PNe  and old PNe. There is no observational value 
for C/H obtained from  CELs, so there is no observational constraint for C/O. 
Following  Allen et al. (1998), we will assume the
following ages for the two PN groups: between 0.1 and 1 Gyr for the young
PN population and between 3 and 9 Gyr for the old one. The present age
of the galaxy is assumed to be 13.5~Gyr. Also in this table we present
the  Fe/H value by Venn et al. (2001) from A-type supergiants.
Since most of the Fe in \ion{H}{ii} regions and PNe is trapped in dust grains 
(Peimbert \& Peimbert 2010; Shields 1978; Delgado-Inglada et al. 2009) 
we do not present the gaseous Fe/H abundance.
In this table we are not showing either He, C and N abundances for PNe because PN
progenitors modify strongly  these element abundances during their evolution, 
thus He, C and N abundances in the PN envelopes are not representative of those
values in the ISM at the formation time of PN progenitors and cannot be
considered as constraints for a chemical evolution model.
Moreover, Cl abundances of PNe are not shown in Table 1, because
they were not determined by HPCG09.


\subsection{The computed models}

We computed four chemical evolution models using different upper-mass
limits for the IMF and different galactic winds prescriptions in order to reproduce
the observed $M_{gas}$ and the O/H values (corrected for dust depletion)
obtained from CELs in \ion{H}{ii} regions. We labeled  them M1C to M4C.
In Table~\ref{modcel}  we present the characteristics and  time evolution
of these four models. In column 1 we list the model name; in column 2,
the upper-mass limit for the IMF used in the model; in column 3, the
corresponding wind prescription: `W' for  well-mixed and `S' for selective
wind; in column 4, the gaseous mass  predicted by the model at the
present time (i.e., 13.5~Gyr); in column 5, we list the time at which the
chemical abundances are obtained, they correspond to the times adopted 
for \ion{H}{ii} regions, young PNe, and old PNe; and in
columns from 6 to 13, the predicted abundance ratios at the different
times. As we mentioned in \S 4.1, in the  three  rows at the bottom of this table,
we include the observed values for \ion{H}{ii} regions, young PNe
and old PNe, which are used as observational constraints.

For model M1C we adopted an $M_{up}$ of 60~M$_\odot$
for the IMF  (equal to the value adopted by CCP06 in their best model), and we
did not include any  outflow. From Table ~\ref{modcel} it is evident that
M1C grossly fails to fit $M_{gas}$ as it predicts a gas mass
$\sim$4.5 times higher than the observed value.

Thus, to reduce the enormous difference between the $M_{gas}$ predicted  
by model M1C and the observed value, we have computed model M2C with a
well-mixed outflow during the early history of the galaxy. 
 This was originally proposed by CCP06  who computed the evolution 
of gaseous thermal energy in the galaxy provided by the SFR, under the assumption 
of no gas outflow (see their Fig. 7). They demonstrated that the accumulated  
thermal energy is much larger than the present binding energy, therefore much of 
that energy should have been thrown away through galactic winds. CCP06 
gave several arguments to explain why the outflows, needed to eliminate this excess in 
thermal energy, likely occurred in the first few billion years of evolution (t $<$ 5  Gyr) when 
the galaxy was less bounded  (note that the binding energy of a galaxy is also 
a function of time and in the hierarchical cosmology, the potential wells of galaxies for a 
given present mass, were shallower in the past). The gas outflow should have stopped when 
the binding and thermal energy came to an equilibrium.

\begin{figure}[ht] \begin{center}
\includegraphics[width=\columnwidth]{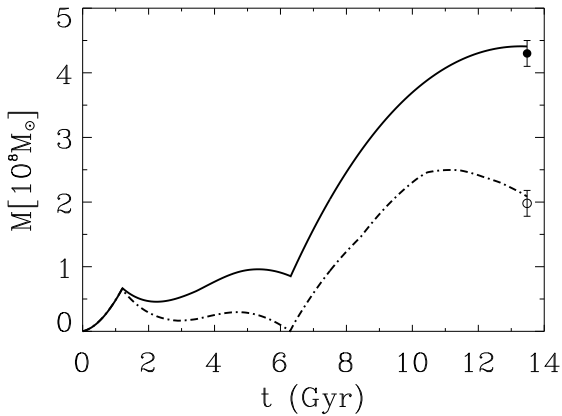} 
\caption{Evolution of the total baryonic mass evolution (solid line) and gaseous mass (dashed line) used
in our  models M2C, M3C, M4C, and M1R.
A well-mixed galactic wind occurs between 1.2 and 6.5 Gyr. Observational constraints are as in CCP06.
 } \label{mgas} \end{center}
\end{figure}

 In our model, the well-mixed galactic wind is turned-on at
$t$=1.2~Gyr, when the star formation starts and it lasts
5.3 Gyr, the time necessary for the ejected mass to be the maximum possible, 
leaving the minimum amount of gas required to maintain the SFR. 
During the wind phase, a value $\nu=6.67$ in the outflow rate equation (Eq.~\ref{windwell}) 
was required. The evolution of the total baryonic mass and the gaseous mass in the galaxy, 
as a function of time, is shown in Fig.~\ref{mgas}. M$_{up}$  These to quantities are directly related with the infall, SFR and galactic wind.
The dashed line is the time evolution of the $M_{gas}$  (Ec.\,1) and the solid line in the time evolution of the total baryonic mass ($dM_{bar}/dt = f(t)-w(t)$).

\smallskip

The present-time O/H value predicted by model M2C is 4 $\sigma$ higher
than the observed values in \ion{H}{ii} regions.  Similarly the ISM O/H ratio
found at $t$=12.9~Gyr is 2 $\sigma$ higher than the observed values of the  
young PN population,  and for the old PN population, the predicted ISM O/H
value at $t$=7.5~Gyr is 2.6 $\sigma$ higher than observed (see Table ~\ref{modcel}) 

Therefore, to better reproduce the O/H abundance ratios we constructed  model M3C,
by adding selective winds to the assumptions of model M2C.
A selective wind, where only the material
produced by MS is expelled to the IGM, might have been caused by the
presence of a notable increase in the star formation rate in the last
600~Myr, as reported by CCP06 and Gouliermis et al. (2010, and references
therein). 
Model M3C required a selective wind between 12.5 and 13.5 Gyr in order to
reproduce  the current O/H value given by \ion{H}{ii} regions. Therefore
this model is identical to  model M2C for t$<$12 Gyr, but reproduces well the observed O/H ratio
and it also reproduces well the O/H values  for both PN populations. For the
other elements which are produced by MS (Ne, Ar, Cl and S) the
 predicted Xi/O ratios are  similar in both models,  unlike those  elements
partially produced by LIMS (C and N), for which the abundance ratios
relative to O are slightly  larger than in M2C  due to the loss of metals
produced by MS in the selective wind phase (see Table~\ref{modcel}). Although
model M3C succeeds in fitting the observed O/H abundances, we do not
consider it a  satisfactory model because the physical mechanism to
produce a 1 Gyr selective wind, where all chemical elements ejected
by all MS are expelled to the IGM before they can mix with the ISM,
is unlikely.  

Then, to fit the O/H ratios observed in \ion{H}{ii}
regions, we proposed a lower value for $M_{up}$, keeping the other
assumptions of  model M2C (a similar model was proposed by HPCG09).
For model M4C we used an $M_{up}=40$ M$_\odot$, including a well-mixed wind,
starting at $t$=1.2~Gyr and lasting  5.3 Gyr.  Figure~\ref{cel2} shows the
evolution  predicted by model M4C for O, C, N, Ne, S, Cl,  Ar and Fe, 
relative to H,  as a function of time. The thick lines (in the eight panels)
represent the model results, the symbols represent the average values observed for
\ion{H}{ii}  regions and the two PN populations as explained in the
figure caption.  We consider model M4C as our best model to reproduce
the chemical abundances derived from CELs, therefore we will discuss
its characteristics in the following subsection.


\subsection{ Our best model for abundances from CELs}

Model M4C was tailored to fit the observed O/H abundance ratio by
reducing  $M_{up}$ to 40~M$_\odot$. We find that it also reproduces well 
the present abundances of the elements heavier than O. This $M_{up}$ may be considered a
low  value for an IMF, however it seems a plausible value since some
authors (e.g., Goodwin \& Pagel 2005 and references therein) have argued
that dwarf galaxies could have smaller $M_{up}$ values than that of the
solar vicinity.

\begin{figure*}[ht] \begin{center} \includegraphics{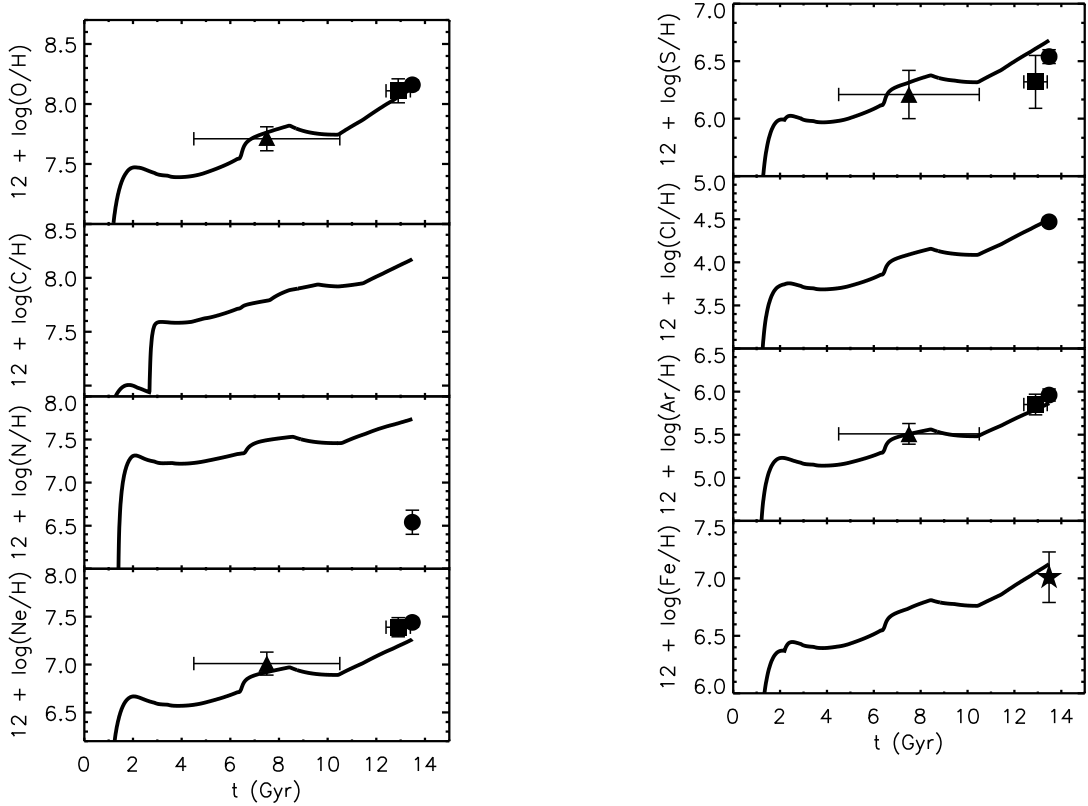}
\caption{Chemical evolution of elements in NGC\,6822 as predicted by model M4C. To
reproduce the observed $M_{gas}$ value and the different heavy element abundances  obtained
from collisionally excited lines for \ion{H}{ii} regions, a well-mixed
galactic wind lasting from 1.2 to 6.5  Gyr and an $M_{up}$=40~M$_\odot$
are assumed in this model. The O/H value of \ion{H}{ii} regions has 
been corrected for dust  depletion, as explained in  \S 4.1. The filled circles, filled
squares, and  filled triangles represent the average observational values for
\ion{H}{ii} regions, young, and old PN populations, respectively. The
filled star in the Fe/H panel represents the value derived by  Venn et al. (2001) 
for  A-type supergiant stars. Observed C/H value is not shown because there is no observed value from CELs.}
\label{cel2} \end{center} \end{figure*}

  From the time evolution shown in Fig.~\ref{cel2}, it is found that for all the elements, the 
temporal enrichment shows two minima, at 4 and 11 Gyr, due to the dilution caused
by the large infall and to the decrease of the SFR (see Fig.~\ref{varfis}),
respectively. We find that the predicted abundances of all the elements  at $t \sim 7$
Gyr agrees with the average abundances of old PNe, whose progenitors were
formed at t = 7.5 $\pm$ 3.0 Gyr (6.0 $\pm$3.0 ~Gyr old). This epoch coincides
with the beginning of the second fast enrichment. Furthermore, the recent
abundances predicted for all the elements matches within uncertainties ($\sim$0.1~dex)  
the average abundances of young PNe, whose progenitors were formed at t =
12.9 $\pm$ 0.5 Gyr  (0.6 $\pm$0.5~Gyr old).  In addition, Fig.~\ref{cel2} 
shows that  model M4C matches well the \ion{H}{ii} region 
average data of all elements within  $\sim$1.5 $\sigma$ (which corresponds to the uncertainties in 
abundance determinations), except in the case of N where it fails grossly; this
problem is discussed in more detail in the next paragraph. Finally, to
fit the observed values of  Fe/H, we needed to assume  a fraction A$_{bin}=0.01$.

 We want to remark that the model fails considerably in reproducing the presently
observed N/H abundance ratio. This is found in several  works on chemical evolution 
models (e.g., Chiappini et al. 2003; Carigi et al. 2005; Moll\'a et al. 2006;  Romano et al. 2010; 
Kobayashi et al. 2011). 
The prediction of our model is 7.4 $\sigma$ higher than the
value observed in \ion{H}{ii} regions, therefore we consider that the N yields for LIMS
adopted, in general,  for the models have been overestimated.

\begin{figure*}[ht] \begin{center} \includegraphics{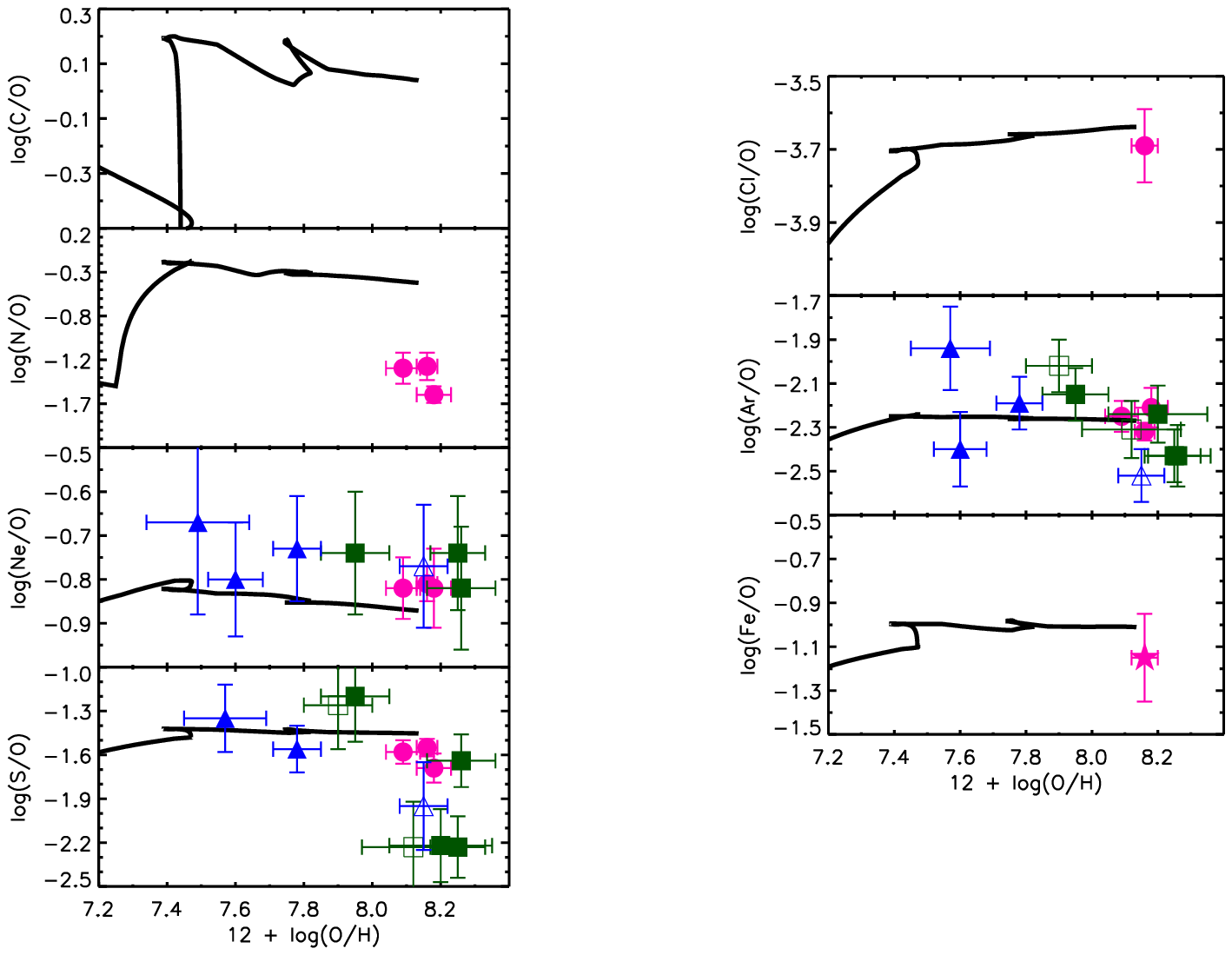}
\caption{Element abundance ratios relative to O as predicted for model M4C compared to observed values of PNe and \ion{H}{ii} regions as presented in HPCG09. The O/H abundance  of \ion{H}{ii} regions has been corrected for dust  depletion, as explained in  \S 4.1. 
\ion{H}{ii} regions are represented by magenta filled circles, young-PNe by green filled squares and old-PNe, by blue filled triangles. The open squares are the two type-I PNe, and the open triangle is PN6 (see text or HPCG09).  The magenta filled star  in the Fe/O panel represents the value derived by  Venn et al. (2001) 
for  A-type supergiants.  Observed C/O value is not shown because there is no observational C/H value from CELs.}
\label{M4c-allC} \end{center} \end{figure*}

   In order to confirm the results given in the previous paragraphs, in Fig.~\ref{M4c-allC} we present
the C/O, N/O, Ne/O, S/O, Cl/O, Ar/O and Fe/O abundance ratios, as a function of 12+log(O/H),  
predicted by model M4C. Here we show the data for the best observed objects published by HPCG09.  
\ion{H}{ii} regions are presented as magenta filled circles, young-PNe are represented by green filled squares and old-PNe, by  blue filled triangles. The open squares are the two  type-I PNe, and the open triangle is PN6 (in  HPCG09 nomenclature) catalogued as an old-PNe based mainly on its Ar/H abundance. We want to remark that  in this figure we present the most complete set of observational constraints for NGC~6822 to be compared with our chemical evolution model.

In this figure, regarding N/O, we find again that the model predicts to much N (probably due to problems with N yields).  Unfortunately, we cannot comment about the C/H abundance because there are no determinations of this element using CELs. 

In the case of Ne, all Ne/O  ratios  in PNe are above  the\ion{H}{ii} values and above model predictions. 
This could be due to the ionization correction factor used to derive Ne  
(Ne/O = Ne$^{++}$/O$^{++}$ is commonly used)  
which is not adequate for low ionization \ion{H}{ii} regions or low density PNe (Peimbert et al. 1992, 1995b).
Ne/O from the model agrees better (within uncertainties) with the values for \ion{H}{ii} regions which means that the Ne and O yields for massive stars up to 40 \msun, assumed in our model, are reliable.  
 
 For S/O, the comparison is not good. 
In particular, the model predictions are 2 $\sigma$ higher than the observed values in young PNe and \ion{H}{ii} regions.  
The uncertainties in S abundance determinations are very large, in particular for PNe, therefore it is difficult to draw any conclusion, regarding S yields.

For Cl/O our model is in  good agreement with the present value of the ISM within the uncertainties. 
Similarly the evolution of Ar/O agrees, within the uncertainties, with the observational values.

We reproduce the present Fe/H  value obtained from  A-type supergiant stars by assuming that only 1\% of all the stars between 3M$_\odot$
 and 15M$_\odot$ become SNIa.

\section{Models for abundances derived from  recombination lines}

It is well known that the C$^{++}$ and O$^{++}$ abundances determined from
the faint \ion{C}{ii} and \ion{O}{ii} recombination lines in photoionized
nebulae are systematically larger than abundances derived from collisionally
excited lines of the same ion (e.g.,  Rola \& Stasinska 1994; Peimbert
et al. 1995; Liu et al. 2004; Esteban et al. 2004; Tsamis et al. 2004;
Wesson et al. 2005; Peimbert \&  Peimbert 2011; Sim\'on-D{\'\i}az \&
Stasi\'nska 2011; Tsamis et al. 2011; Carigi \&  Peimbert 2011). Other ionic abundances such as O$^+$/H$^+$ and N$^+$/H$^+$, 
derived from recombination lines show the same behavior. This 
problem is known as the "Abundance Discrepancy" and is generally 
parametrized by the abundance discrepancy factor (ADF) defined as: 

\begin{equation} \label{adf} {\rm
ADF(X^{i+})=( X^{i+}/H^+)_{RL}/(X^{i+}/H^+)_{CEL}}.  \end{equation}

\noindent ADFs have commonly values of about 2. Then, trying  to
discriminate between the abundance ratios obtained from both, CELs and
RLs, we have computed a model to reproduce the O/H abundance calculated from
RLs and we compared it with our best model based on collisionally excited
lines (model M4C).


\subsection{Observational constraints}

 We used the abundances reported by Peimbert et al.  (2005), from RLs,  for   the \ion{H}{ii}
region HV.  They obtained a value  12 + log(O/H)=8.42$\pm$0.06, where this determination 
includes the correction for dust depletion. Peimbert et al.  (2005) considered  dust corrections 
of 0.08 dex  for  O and 0.10 for  C, these corrections have been adopted for \ion{H}{ii} regions throughout this paper. Abundance values for other elements are taken from Table 9 of Peimbert et al. (2005).

On the other hand, there is no  abundance determinations  based on RLs
for  PNe in NGC 6822 because  these faint lines  were not detected in
these objects. Thus we  have not  direct observational constraints based
on RLs for the past component of  the ISM.  Therefore we have  used the
determinations of chemical abundances derived from CELs as reported by
HPCG09 and we have modified them by considering a certain constant value 
for the ADF. Due to this procedure, the RLs abundances for PNe should be considered only indicative.

Liu et al.  (2006) report that for most galactic PNe,  ADFs are in
the range of 1.6 to 3. Likewise Peimbert et al. (1995a) determined
ADFs for many galactic PNe and computed similar ADF values.
Therefore we have applied an ADF=2 to the PN abundance determinations
from CELs  to derive a value for recombination lines. Note that the ratio
of any heavy element to O in PNe,  is not affected  by the ADF correction 
because the abundances of both elements have been increased by the same amount.

In the bottom part of Table ~\ref{modrel} we present the chemical
abundances for \ion{H}{ii} regions obtained via recombination lines
plus correction due to dust depletion. In addition we present  the
chemical abundances for the young and old PN populations calculated after 
using the ADF mentioned above for the abundance values by HPCG09.


\subsection{The model}

We computed a single chemical evolution model using similar assumptions
to those described for model M4C. The model is labeled  M1R and it was
tailored to reproduce the O/H  abundance obtained for the \ion{H}{ii}
region HV, from recombination lines.

The same as in model M4C, model M1R required a well-mixed wind, starting at 1.2 Gyr and lasting
5.3~Gyr,  to reproduce $M_{gas}$. Now an  $M_{up}$=80 M$_\odot$ is needed to
reproduce the current O/H value. In Table~\ref{modrel} (top part) we
present the gaseous mass and the O/H, C/O, N/O, Ne/O, S/O, Cl/O, Ar/O,
and Fe/H ratios predicted by model M1R for different times.

In this model, the computed O/H value at $t = 7.5$ Gyr is in very good agreement
with the adopted value for the old PN population. For the young PN population, the oxygen
abundance calculated, is also in very good agreement with the  adopted value 
which is very similar to the one in \ion{H}{ii} regions.
To fit the observed Fe/H value we required to adopt, again, A$_{bin}$ = 0.01,

\begin{figure*}[ht] \begin{center} \includegraphics{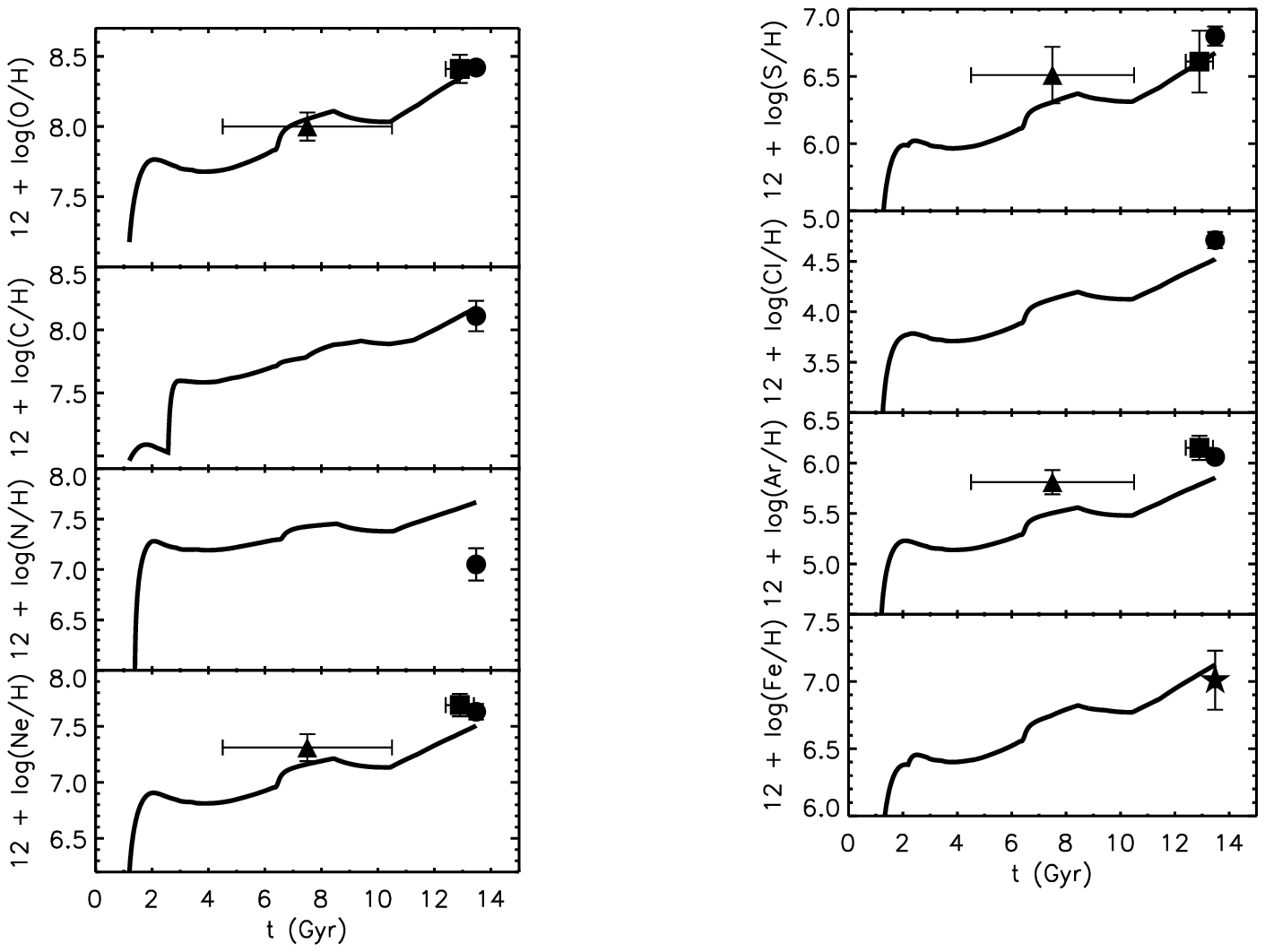}
\caption{Chemical evolution of elements in NGC\,6822 as predicted by model M1R. To
reproduce the observed $M_{gas}$ value and the different heavy element abundance ratios  obtained
from RLs for \ion{H}{ii} regions, a well-mixed
galactic wind lasting from 1.2 to 6.5 Gyr and an $M_{up}$=80~M$_\odot$
are assumed in this model. The observed O/H and C/H values have been corrected
for dust  depletion, as explained in  \S 4.1. The filled circles, filled
squares, and  filled triangles represent the average observational values for
\ion{H}{ii} regions, young, and old PN populations, respectively. The
filled star in the Fe/H panel represents the value derived by  Venn et al. (2001) 
for  A-type supergiant stars.
}
\label{xivstrl} \end{center} \end{figure*}

 Fig.~\ref{xivstrl} shows the time evolution of the element abundances (O, C, N, Ne, S, Cl, Ar, and Fe) 
relative to H as  
predicted by model M1R. The thick lines in every panel represent our model 
results and the symbols are equal to those in Fig. 3.  
In this model the computed C/H value agrees well 
with the observational value obtained from RLs. The predicted N/H
ratio is slightly closer to the observational constraint than in the model M4C, but it is still
3.6 $\sigma$ higher than the observed value, again indicating that
the N yields for LIMS have to be revised. The predicted evolution of Ne/H shows relatively good agreement with observed values of old PNe but not so good for young PNe and \ion{H}{ii} regions. In the case of S/H, the predicted evolution is in a good agreement 
with the PNe observed values but not with observations of \ion{H}{ii} regions. 
The predictions for Cl/H and Ar/H are significantly lower than the observed values, in all the cases.  

\begin{figure*}[ht] \begin{center} \includegraphics{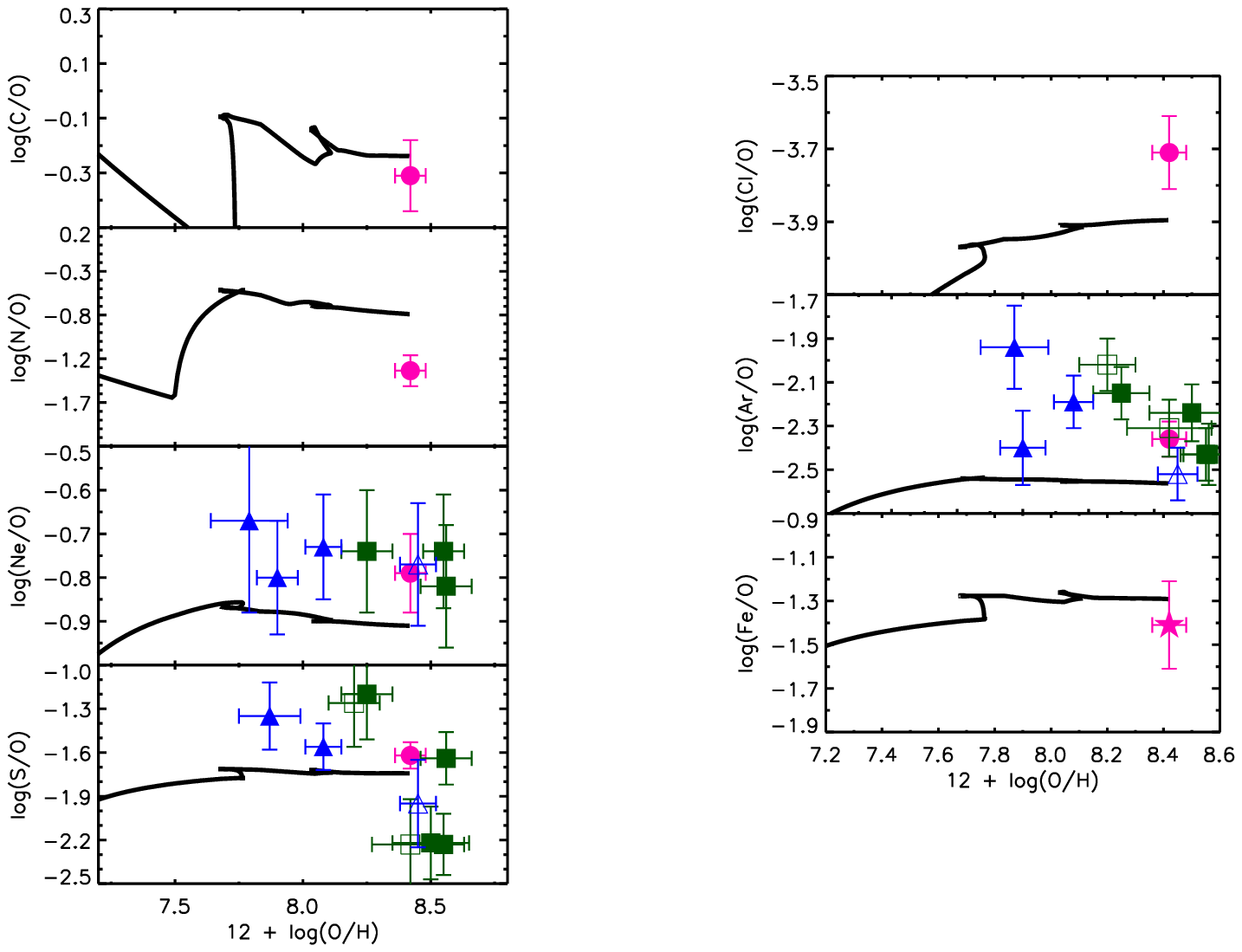}
\caption{Element abundance ratios, relative to O, as predicted for model M1R compared to observed values from recombination lines for the \ion{H}{ii} region HV  as presented in Peimbert et al. (2005) and the adopted values for PNe, as explained in \S 5.1. O and C abundances in HV have been corrected
for dust  depletion, as explained in  \S 4.1. \ion{H}{ii} region HV is presented in magenta filled circl, young-PNe are represented by green filled squares and  old-PNe, by blue filled triangles. The open squares are the two type-I PNe  and the open triangle is PN6 (see HPCG09).  The
magenta filled star, in the Fe/O panel, represents the value derived by  Venn et al. (2001) 
for  A-type supergiant stars.}
\label{xivsorls} \end{center} \end{figure*}

The described behaviors are clearly noticed in Fig.~\ref{xivsorls} where we present the predicted evolution of the   C/O, N/O, Ne/O, S/O, Cl/O, Ar/O and Fe/O  abundance ratios as a function of 12+log O/H and the individual values of the best observed objects (see \S 4.3). In this figure, many of the Ne/O observed values (in particular those of young PNe and \ion{H}{ii} regions) lie above the predictions, 
the S/O for \ion{H}{ii} regions is well predicted within 1$\sigma$ level. The predicted  Cl/O and Ar/O values are $\sim$ 2 $\sigma$ lower than the observed values. 
This latter problem is due to the flat extrapolation assumed for the SN yields for $m  > 40 $ \msun ~(see \S3).
According to Woosley \& Weaver (1995),  Cl and Ar yields for $m  = 40 $ \msun \ are lower than for $m = 35 $ \msun \
and therefore very massive stars would contribute less than massive stars to the ISM enrichment of those elements.

 We consider that to make a better comparison between models and observations SN yields for  $m  > 40 $ \msun \ are necessary. Recall that the yields 
in this paper for $m  > 40 $ \msun \ were extrapolated from the $m = 40 $ \msun \ yields 
and this extrapolation seems not adequate.

 Finally, to fit the observed log [Fe/H]=$-$0.49 value  we adopted A$_{bin}$=0.01, the same value used for CELs models.

 \subsection{The stellar yields problem}

 One of the most important elements in the chemical evolution models are the stellar yields. 
In the literature there are several works about 
 stellar evolution models that present stellar yields computed from different codes.
These stellar models include different physical ingredients and it is difficult to compare them
(e.g., Romano et al. 2010 and references therein).
In addition, there is no complete grid of stellar yields available
for the entire range of stellar masses.
 
Our main problem to reproduce RLs observations for some chemical species is the lack of a 
homogeneous and complete set of stellar yields. 
First of all, as we explained in detail in \S3, for LIMS we use the Karakas \& Lattanzio (2007) yields, 
which cover from 1 to 6 M$_\odot$ and for MS,  those from  the Geneva group (from 9 to 80 \msun) for the pre-SN stage plus those from 
Woosley \& Weaver (1995, hereafter WW) (for 9 to 40 \msun) for the SN stage, together with the special treatment proposed by Carigi \& Hernandez (2008).
In order to reproduce the abundances  from RLs, we need to extrapolate the WW yields from 40 M$_\odot$ to 80 M$_\odot$. 
We can see in Figs. 3 and  5 that the predictions for the evolution of O/H and Ne/H are in good agreement with the observational constraints obtained from CELs (model M4C) and from RLs (model M1R) 
for the present (\ion{H}{ii} regions) and past (PNe) components of the ISM.
However for the other chemical elements, C/H, N/H, S/H, Cl/H, Ar/H, and Fe/H the predictions do not change for the two models, even when for model M1R M$_{up}$ is 80 \msun. 

Moreover, Cl/H and Ar/H values predicted by  model M4C are in  good agreement with observations, but this 
is not the case for model M1R. 
We blame this to the lack of SN yields for masses $>$ 40 \msun and the extrapolation made by us up to  80 M$_\odot$. 
We have to keep in mind that M$_{up}$ of the model  M4C is equal  to 40 M$_\odot$ and no extrapolation for the yields was needed, while  M$_{up}$ of M1R is equal to 80 M$_\odot$ and we have to extrapolate the yields. 

 \begin{figure}[ht] \begin{center} \includegraphics[width=\columnwidth]{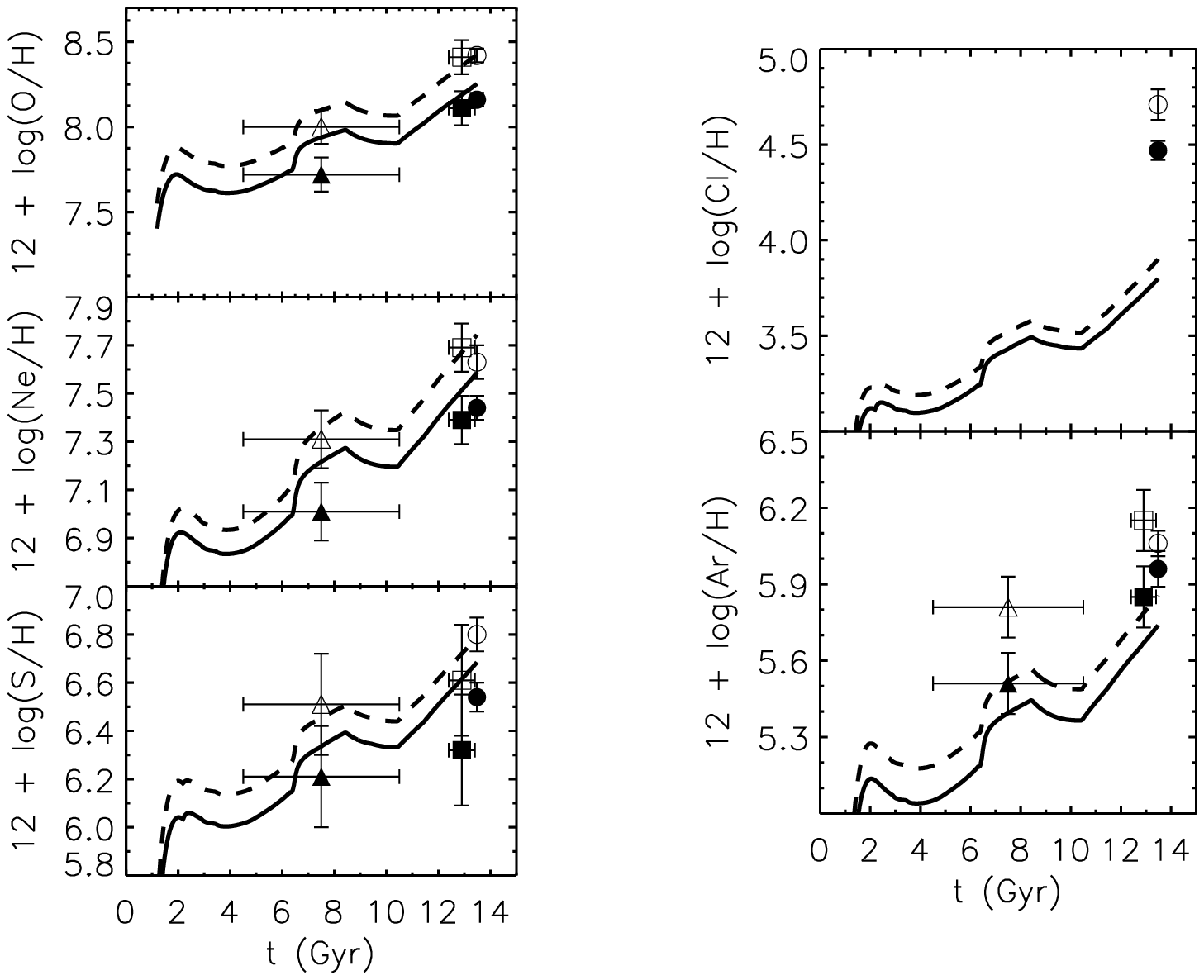}
\caption{We present the evolution for O/H, Ne/H, S/H Cl/H and Ar/H for two models, M4Ckoba (solid line) and M1Rkoba (dashed line), based on Kobayashi et al. (2006) yields. The filled circles, filled squares, and  filled triangles represent the average observational values for
\ion{H}{ii} regions, young, and old PN populations, respectively.  }
\label{xivstimekoba} \end{center} \end{figure}

Once we have these results, we did an exercise changing only the stellar yields of massive stars.
We took the SN yields by 
Kobayashi et al. (2006) and the pre-SN yields by the Geneva's group. 
We kept the rest of assumptions described in \S3 and constructed two models, 
M4Ckoba (solid line) and M1Rkoba (dashed line) presented in Fig~\ref{xivstimekoba}.

We plot the evolution of the O/H, Ne/H, S/H, Cl/H, and Ar/H in order to compare them with the observations.
The predictions for the evolution of  Ne/H and S/H abundances are in very good agreement  with the observational constraints 
obtained from RLs only (dashed line and open symbols) but for CELs (solid line and filled symbols) 
the model predictions are $\sim 3 \sigma$ higher. 
Based on O/H, Ne/H and S/H, Kobayashi et al. (2006) yields produce a good model for observed RLs abundances but not for observed CELs abundances.
If we want to force the M4Ckoba model to reproduce abundances from CELs, we must diminish even more  M$_{up}$ in the IMF (M4C has an M$_{up}$=40\,M$_\odot$). 
The Cl/H and Ar/H predictions of both models are lower than the observational constraints,  by $\sim 0.75$ and $\sim 0.25$ dex, respectively.
In order to reproduce the Cl/H and Ar/H observed abundances in \ion{H}{ii} regions, the Cl and Ar yields should be 5.6 and 1.8 times higher than those computed by Kobayashi et al. (2006).
This is an important result, because previous works that test stellar yields (e.g.,
Timmes et al 1995; Romano et al. 2010; Kobayashi et al. 2011)
cannot check the Cl and Ar yields due to the lack of abundance determinations of these elements  in stars of the solar vicinity.

Based on this exercise, we prefer to keep the SN yields by WW for the model reproducing abundances from CELs (M$_{up}=40$ \msun).
Nevertheless,  we prefer the SN yields by Kobayashi et al. (2006)  for the model reproducing RLs abundances (M$_{up}=80$ \msun)
with Cl and Ar yields increased by factors of $\sim 6$ and $\sim 2$, respectively.


\section{The $\Delta${\it Y}/$\Delta${\it O} ratio and the IMF}

Our chemical models predict the evolution of the He and O enrichment, both
by mass,  known as $\Delta${\it Y}  and $\Delta${\it O},  as a function of
time. For convenience, and to compare with observations the $\Delta${\it
Y} /$\Delta${\it O}  ratio is often used, where $\Delta${\it O} is
equal to {\it O} and $\Delta${\it Y} is equal to ${\it Y- Y_p}$. Here
${\it Y_p}$ is the primordial He abundance and amounts to 0.248, the
value derived from metal poor irregular galaxies and WMAP (Peimbert et
al. 2007a;  Dunkley et al. 2009). It is well known that $\Delta${\it Y}
versus $\Delta${\it O} mainly depends on the adopted yields and the IMF.

\begin{figure}[ht] \begin{center} 
\includegraphics[width=\columnwidth]{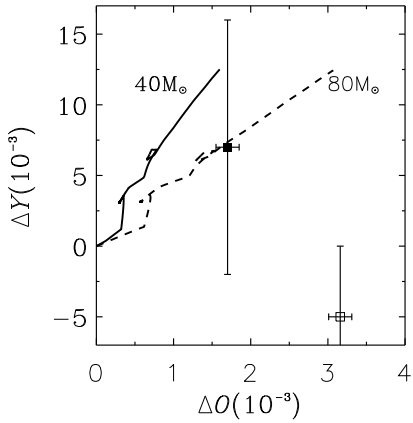} \caption{  Evolution
of $\Delta${\it Y} vs. $\Delta${\it O} for our two best models, M4C
(solid line) and M1R (dashed line).  The black square represents observed
abundances under the assumption of constant temperature, and the open square represents 
abundances from RLs under the assumption of temperature variations (Peimbert et al. 2005), 
note that the observed  $\Delta${\it Y}  values might be lower limits due to the possible 
presence of neutral helium inside the \ion{H}{ii} region. } 
\label{YvsO} \end{center} \end{figure}

Fig.~\ref{YvsO} shows the $\Delta${\it Y} vs. $\Delta${\it O} behavior for our
models M4C (solid line, this model reproduces the O/H abundances given
by CELs) and M1R (dashed line, reproducing the O/H abundances given by
RLs ). Both lines show a similar behavior (same form, different slope)
which depends mainly on the history of the mass accretion, the variation
of the SFR with time, and on the time delays assigned to the LIMS for
their contribution to the chemical enrichment of He.

In Fig.~\ref{YvsO}  the initial slope corresponds to the first epochs of the galaxy,
where only infall and IRA from MS are the processes controlling the He
and O abundances. The sudden discontinuity with a large increase in He,
occurs when LIMS start their contribution to the He enrichment of the ISM;
these stars mainly produce He but not O. From this point onwards both
curves evolve in a continuous increase of He and O, with an almost constant
slope, perturbed only by a small loop which occurs when the SFR changes
suddenly, in the epoch  from 6.3 to 10.5 Gyr (see Fig.~\ref{varfis} bottom). At
even later times both curves increase with an apparently  constant slope.

Interestingly, although the behavior is similar,  both curves showing the
same perturbations due to the behavior of the SFR, the slopes are very
different. As expected, the slope is larger for M4C, which has $M_{up}$ of
40 M$_\odot$, and thus in this model less oxygen is produced. The slope
derived for model M4C, considering only the zone from $\Delta {\it O}
> 0.5 \times 10^{-3}$,  is about 7.2, while for  model  M1R it is about 3.7.

In Fig.~\ref{YvsO}  we also include two observational points. The filled square
represents the ($\Delta${\it Y}, $\Delta${\it O}) values  derived from the \ion{He}{i} RLs  by Peimbert et al. (2005)
and the [\ion{O}{iii}] CEL (HPCG09) under the assumption of a constant temperature given by the 
[\ion{O}{iii}] CELs. The open square represents the values derived
from abundances obtained through \ion{He}{i} and \ion{O}{ii} RLs, considering the presence of spatial temperature
variations. It is observed that, in both cases, $\Delta${\it O} is well fitted because the
models were tailored to fit O/H.  On the other hand, the observed $\Delta${\it
Y} shows  huge uncertainties, because it is very difficult to derive a precise
value for He abundances due to: i) the large statistical errors due to
the faintness of the lines (represented in the Fig.~\ref{YvsO}  by the error bars),
and ii) the possible presence of neutral He in the observed \ion{H}{ii}
regions, that would increase the derived $\Delta${\it Y} value for both
determinations. Both {\it Y} values are derived from RLs, one under the assumption of
constant temperature and the other under the assumption of the presence of temperature
variations, the second one gives a {\it Y} value smaller than the primordial value
probably indicating the presence of neutral helium inside the observed \ion{H}{ii} region.

We can compare also the slopes predicted by our models with observational
data in the literature.  For instance, Izotov et al. (2007) have derived
chemical abundances  for a large number of \ion{H}{ii} galaxies of
different metallicities. From these data they derived a slope $\Delta${\it
Y}/$\Delta${\it O} =  3.6$\pm$0.7 or 3.3 $\pm$0.6, depending on the
\ion{He}{i} emissivities used, these numbers have not been corrected by
the fraction of O atoms trapped in dust grains. By considering that such dust fraction 
 amounts to 0.08 dex (Peimbert et al. 2005)  we obtain from Izotov et al. (2007) data that  $\Delta${\it
Y}/$\Delta${\it O} =  3.0$\pm$0.6 or 2.75 $\pm$0.5.  On the other hand,
Peimbert et al. (2007b), from theoretical and observational results,
obtain $\Delta${\it Y}/$\Delta${\it O} = 3.3$\pm$0.7. These
slopes are in better agreement with model M1R, which was computed to fit
O abundances from RL and for which an $M_{up}$= 80 M$_\odot$ was assumed.

To summarize, model M1R produces a $\Delta${\it Y}/$\Delta${\it O}
slope similar to the one derived for a set of other galaxies, while
M4C does not. In the case of NGC\,6822, however, a much better {\it Y}
determination  is needed to discuss this problem further.


\section{Discussion and conclusions}

We have developed a code to calculate chemical evolution models for
galaxies. The code includes the instant recycling approximation for MS,
and a time delay  for the contribution of LIMS. Such a time delay
has been obtained for each element following the prescription given
in Franco \& Carigi (2008). Our code also includes the SNIa contribution to the chemical
evolution in the galaxy. The code can be used for different SFRs, IMFs,
yields, infalls and outflows. It predicts the evolution of 27 elements. 

Using this code we calculated chemical evolution models for
the dwarf irregular NGC\,6822. The main characteristics of the models for this galaxy, are:

\noindent - A time dependent infall, parametrized from the one obtained by CCP06 from cosmological
 considerations, with a primordial abundance given by {\it X}=0.752, {\it Y}=0.248 and {\it Z}=0.000 
 (Fig.~\ref{varfis}).

\noindent - The star formation history derived by CCP06, which is based on the
photometric properties of the galaxy (Fig~\ref{varfis}).

\noindent - The IMF by Kroupa et al. (1993) with different M$_{up}$ values depending on
the model.

\noindent - A set of metallicity dependent yields for 27 elements
including  H, He, C, N, O, Ne,  S, Cl, Ar, Fe and others.

\noindent - 1\% of the stars with masses between 3 and 15 M$_\odot$ gives
up binary systems and every one of these systems becomes a SNIa. The
contribution of SNIa enrich the ISM with a time delay of 1~Gyr after
the formation of the SNIa precursor.

\noindent - Two types of time dependent outflows can be used: well mixed and
selective.  

\medskip

Our models for NGC 6822 were constructed to fit the following
observational constraints: a) $M_{gas}$, b) O/H chemical abundances of \ion{H}{ii} regions and the
Fe/H ratio of A-type supergiant stars representing the current abundances
of the ISM, and c) chemical abundances of young (0.6$\pm 0.5$ Gyr) and old (6$\pm 3$ Gyr) planetary
nebulae representing the abundances of the ISM at the time they were
formed. The \ion{H}{ii} abundances have been corrected for the fraction
of C and O in dust grains. Two sets of chemical abundances were used: 
those derived from CELs (HPCG09) and those derived from RLs
(Peimbert et al. 2005), they differ by  an abundance
discrepancy factor, ADF, of about 0.26 dex.
 As the RLs were not observed in PNe, we estimated their RL abundances based on the CEL abundances, and adopting an ADF of 2.  Thus PNe abundances from RLs are only indicative.

The baryonic total mass captured by the system is given by the gas
inflow estimated from cosmological considerations, and the amount of gas
transformed into stellar mass is obtained from photometric data. From
these two considerations only, the remaining gas turns out to be 4.5 times
higher than observed. Therefore to reproduce the observed $M_{gas}$
value the models required a well mixed outflow lasting from  $t=1.2$~Gyr (which
is the time when the star formation starts) to $t = 6.5$ Gyr  (when the gas mass has diminished
enough to fit the present gaseous mass but without interrupting the star formation, Fig.~\ref{mgas}).

 The SFR used in our models was derived from photometric data  by CCP06, and it is in good agreement with the abundances obtained for old PNe (HPCG09). These two studies are independent and therefore old PNe data support the SFR based on photometric data.

The two best models obtained to reproduce the present day ISM O/H value
required: a) an $M_{up}=40$ \msun \ for model M4C, which is based on
 abundances derived from CELs,  and b) an $M_{up}=80$ \msun \ for model M1R,
which is based on  abundances derived from RLs.

Both models reproduce the O/H evolution of the past component of the ISM
obtained from young and old PNe.  This is particularly true for model M4C based on CELs. 
This result implies that the adopted SFR, derived from photometric data,  describes well  
the shape of the O/H evolution and that the inferred well mixed
wind is consistent with the absolute O/H value.

The O/H ratio depends strongly on the use of RLs or CELs, but since the models are tailored 
to fit the observed values, it is not possible to decide which abundances are more 
adequate  based on O/H alone.

 We have tested the time evolution of other elements (C, N, Ne, S, Cl, Ar and Fe, relative to H).
In general, the predictions of model M4C agree with the observations of \ion{H}{ii} regions and old and 
young PNe, indicating that the assumptions used for this model are correct and that the SN yields used in the 9 to 40 M$_\odot$ range, derived  from WW, are adequate.
\medskip

About 50\% of  C is produced by LIMS and about 50\% by MS in the pre-SN phase. The C/O ratio is 
well fitted by model M1R, result that implies that the C and O yields are adequate.

On the other hand the time evolution of heavy elements other than O, predicted by model M1R, 
are not always in agreement with the observations. Given that the predicted abundance ratios 
depend strongly  on the yields,  the adopted yields can then be tested. The C/O, S/O and Fe/O ratios predicted   are  in fair agreement with the observations.  But for Ne/O, Cl/O and Ar/O,
the predicted values are smaller than observed, probably indicating that
the extrapolation of the SN yields from 40 to 80 M$_\odot$ made in this
paper is not good enough. A different set of yields for this mass range, provided by Kobayashi et al. (2006) was tested,  finding a good agreement with observational values in both models (M4Ckoba 
and M1Rkoba, Fig~\ref{xivstimekoba}) for Ne/H and S/H,  but Cl/H and Ar/H are underestimated by factors of $\sim$6 and $\sim$2 respectively. It is worth  to mention that this is the first time that observational constraints for these elements are presented and compared with the models in order to test the stellar yields.

Model M1Rkoba is in very good agreement with the observed value of O/H obtained from RLs, but
 in order to reproduce the observed value obtained from CELs (model M4Ckoba) it is necessary to lower M$_{up}$ below 40 $\msun$. We require a new set of yields for the 40 to 80 M$_\odot$ mass range to make a better test of the model  M1R  against observations.

The predicted N/H values in both models are always considerably higher than observed and
probably indicate that the N yields need improvement. About 80\% of N is
produced by LIMS and about 20\% by MS in the pre-SN stage.  Therefore
lower N yields for LIMS would provide a better fit to the observed N/O values.   

At this point, we are not able to conclude which set of abundances (based on CELs or RLs) represents 
better the real abundances in the ISM.
Model M4C , that assumes $M_{up}=40$ \msun, reproduces well
the  O/H, Ne/H, S/H, Ar/H, Cl/H, and Fe/H observational constraints.
While model M1R, that assumes $M_{up}=80$ \msun, reproduces well
the  O/H, C/H, Ne/H, S/H, and Fe/H observational constraints,
but fails to reproduce Cl/H and Ar/H 
probably due to inadequate SN yields for stars with masses above 40 \msun.
\smallskip

Since $\Delta Y$ does not depend strongly on the adopted temperature
and $\Delta O$ does, $\Delta Y/\Delta O$ is a very good way to
discriminate between M1R and M4C models. Unfortunately the helium
abundance determination for NGC 6822 is poor due to the errors in
the He I line intensities and the correction due to the neutral He
inside the \ion{H}{ii} regions.  We have used the $\Delta Y/\Delta O$
determinations derived by other groups, based on other galaxies, and we
find that the M1R model agrees better with the observations of the other
galaxies than the M4C model.
A much more precise He abundance determination of the \ion{H}{ii}
regions in NGC 6822 is needed to advance in this topic.

\begin{acknowledgements} 
We thank  the referee Dr.  G. Lanfranchi for his comments and suggestions which helped to largely improve this paper. 
We are grateful to F\'atima Robles-Valdez for useful discussions.
M. Pe\~na is grateful to DAS, Universidad
de Chile, for hospitality during a sabbatical stay when part of this
work was performed.  L. H.-M. benefited from the hospitality of the
DAS, Universidad de Chile for this work.
L. H.-M. received a scholarship from  CONACYT-M\'exico and DGAPA-UNAM.
M. Pe\~na gratefully acknowledges financial support from FONDAP-Chile and
DGAPA-UNAM. This work received financial support from CONACYT-M\'exico
(grants  60354, 129753) and   DGAPA-UNAM (grants  IN-112708 and IN-105511).
\end{acknowledgements}



\begin{landscape} 
\begin{table}[t] 
\begin{center}
{\footnotesize 
\caption{Models that reproduce abundances calculated via
collisional excited lines. The `W' and `S' are for well-mixed and
 selective wind prescription, respectively. } 
\label{modcel}
\begin{tabular}{lcccccccccccc} 
\hline 
\hline 
Model & $M_{up}$ & wind & $M_{gas}$ & $t$ & O/H$^a$ & log(C/O) & log(N/O) & log(Ne/O)& log(S/O) & log(Cl/O) & log(Ar/O) &log(Fe/O) \\
           & (M$_\odot$) &     & (10$^8$M$_\odot$) & (Gyr)& & & & & & & & \\ 

\hline
M1C   &  60            & -         & 8.99& 13.50 & 7.95 & -0.08 & -0.44 & -0.85 & -1.63 & -3.87 & -2.45 & -1.20\\
      &     &  &               & 12.90 & 7.90 & -0.08 & -0.43 & -0.85 & -1.63 & -3.87 & -2.45 & -1.20\\
      &     &  &                   &  7.50 & 7.68 & -0.09 & -0.39 & -0.84 & -1.63 & -3.89 & -2.44 & -1.21\\
\hline
M2C   &   60 & W& 2.08  &13.50 &8.32   & -0.15 & -0.63 & -0.88 & -1.65 & -3.81 & -2.47 & -1.20\\
      &      &  &                   &12.90 &8.25   & -0.14 & -0.61 & -0.88 & -1.65 & -3.81 & -2.46 & -1.20\\
      &      &  &                      & 7.50  &7.96   & -0.17 & -0.50 & -0.87 & -1.65 & -3.84 & -2.45 & -1.21\\
\hline
M3C  &   60&W-S&2.08 &13.50 & 8.19 & -0.06 & -0.52 & -0.88 & -1.63 & -3.81 & -2.46 & -1.17\\
     &     &   &                   & 12.90& 8.20 & -0.14 & -0.60 & -0.88 &-1.65  & -3.82 & -2.46 & -1.20\\
     &     &   &                      & 7.50  & 7.96  & -0.17 & -0.50 & -0.87 & -1.65 & -3.84 & -2.45 & -1.21\\
\hline
M4C   & 40& W  & 2.08& 13.50 & 8.13  &  0.04  & -0.40  &  -0.87  & -1.45   & -3.64  &  -2.27  &  -1.00\\
          &    &      &            &12.90& 8.05  &  0.05  & -0.37   & -0.87  & -1.45   & -3.64  &  -2.27  &  -1.01\\
          &    &      &             &  7.50 & 7.76   &  0.02 & -0.27   &  -0.84  & -1.43  &  -3.67  &  -2.26  &  -1.00\\
   \hline
\multicolumn{4}{l}{Observational constraints} \\
   \hline
\ion{H}{ii} region, A stars&\tiny    & \tiny   & \tiny 1.98$^b$$\pm$0.20        & \tiny 13.50              & \tiny 8.16$^c$$\pm$0.04   &           & \tiny -1.61$\pm$0.14   &  \tiny -0.72$\pm$0.06     & \tiny -1.61$\pm$0.10      & \tiny -3.69$^d$$\pm$0.06    & \tiny -2.20$\pm$0.08   &\tiny -1.15$^e$$\pm$0.20\\
   PN young                                 & \tiny   & \tiny   &\tiny                             &\tiny 12.90$\pm$0.50  & \tiny 8.11$\pm$0.10                     &\tiny                                                 & \tiny                               &  \tiny -0.77$\pm$0.14     & \tiny  -1.80$\pm$0.20     & \tiny                                          & \tiny -2.26$\pm$0.16   & \tiny \\     
   PN old                                      &\tiny    & \tiny   &  \tiny                           & \tiny 7.50$\pm$3.00   & \tiny 7.72$\pm$0.10                     &  \tiny                                               & \tiny                               &  \tiny -0.74$\pm$0.16     & \tiny  -1.62$\pm$0.20     & \tiny                                          & \tiny -2.26$\pm$0.16   &  \tiny \\
   \hline
\multicolumn{4}{l}{$^a$ \tiny Abundances in units of 12 + log(O/H).} \\
\multicolumn{4}{l}{$^b$ \tiny CCP06.}\\
\multicolumn{5}{l}{$^c$ \tiny Adding 0.08 dex  to O  gaseous value due to depletion in dust grains.} \\
\multicolumn{4}{l}{$^d$ \tiny Peimbert, Peimbert \& Ruiz (2005).} \\
\multicolumn{5}{l}{$^e$ \tiny Fe/H from A-stars by Venn et al. (2001) and CEL O/H from \ion{H}{ii} regions.} \\
\hline \hline
\end{tabular}
}
\end{center}
\end{table}
\end{landscape}


\begin{landscape}
\begin{table}[t]
\begin{center}
{\footnotesize
\caption{Model that reproduces abundances calculated via recombination lines plus correction for dust depletion. 
In column 3, `W' is for well mixed wind prescription. }
\label{modrel}
\begin{tabular}{lcccccccccccc}
\hline
\hline
Model  & $M_{up}$&wind& $M_{gas}$ & $t$ & O/H$^a$ & log(C/O) & log(N/O) & log(Ne/O) & log(S/O) & log(Cl/O) & log(Ar/O) & log(Fe/O) \\
             &(M$_\odot$)&      &(10$^8$M$_\odot$)&(Gyr)&         &          &         &         &           &         &          &      \\
\hline
M1R       & 80  &  W  & 2.08 & 13.50      & 8.41 & -0.24 & -0.75&  -0.91 & -1.74&  -3.89 & -2.56  &-1.29\\
          &     &     &                  & 12.90      & 8.34 & -0.24 & -0.74 & -0.91 & -1.74&  -3.90 & -2.56  &-1.29\\
            &     &     &               &  7.50       &  8.05 & -0.27& -0.63&  -0.89 & -1.74       &  -3.93 & -2.55  &-1.30\\
 \hline
\multicolumn{4}{l}{Observational constraints}  \\
   \hline
  \tiny \ion{H}{ii} regions$^c$, A stars      &\tiny    & \tiny     & \tiny 1.98$^b$$\pm$0.20       & \tiny 13.50                  & \tiny 8.42$\pm$0.06 &  \tiny -0.31$\pm$0.13 & \tiny-1.37$\pm$0.17  &\tiny -0.79$\pm$0.09      &\tiny -1.62$\pm$0.09     &\tiny -3.71$\pm$0.10    & \tiny -2.36$\pm$0.08 &\tiny-1.41$^d$$\pm$0.20\\
   PN young                                             & \tiny   & \tiny   &\tiny                                           &\tiny 12.90$\pm$0.50        & \tiny 8.41$\pm$0.10 &   \tiny                               & \tiny                               &  \tiny -0.77$\pm$0.14     & \tiny  -1.80$\pm$0.20     & \tiny                             & \tiny -2.26$\pm$0.16 & \tiny \\  
   PN old                                                  &\tiny    & \tiny   &  \tiny                                         & \tiny 7.50$\pm$3.00          & \tiny 8.02$\pm$0.10 &  \tiny                                & \tiny                               &  \tiny -0.74$\pm$0.16     & \tiny  -1.62$\pm$0.20     & \tiny                            & \tiny -2.26$\pm$0.16 &  \tiny \\
   \hline
\multicolumn{4}{l}{$^a$ \tiny Abundances in units of 12 + log(O/H).} \\
\multicolumn{4}{l}{$^b$ \tiny CCP06.}\\
\multicolumn{4}{l}{$^c$ \tiny Peimbert, Peimbert, \& Ruiz (2005).} \\
\multicolumn{4}{l}{$^e$ \tiny Fe/H from A-stars by Venn et al. (2001) and RL O/H from \ion{H}{ii} regions.} \\
\hline \hline
\end{tabular}
}
\end{center}
\end{table}
\end{landscape}


\newpage

\section{Appendix A}
In this section we present the returned masses, time delays,
and integrated yields for the chemical elements considered in this work.

In Table 3 we show the time delays of LIMS for the returned mass and for the integrated yields of
eight chemical elements. These data are independent of the $M_{up}$ value adopted in the IMF,
but they are dependent on the initial stellar metallicity $Z_i$.

The data shown in Tables 4 and 5 were obtained by assuming  $M_{up}=40$ \msun \, and $M_{up}=80$ \msun, respectively.


\begin{table*}
\begin{center}
\caption{Time delay (Gyr) of LIMS.}
\label{taus40}
\begin{tabular}{cccccccccc}
\hline
\hline
$Z_i$ & $\tau_m$ & $\tau_H$ & $\tau_{He}$ & $\tau_C$ & $\tau_N$ & $\tau_O$ & $\tau_{Ne}$ & $\tau_S$ & $\tau_{Fe}$ \\
\hline
1.0$\times10^{-5}$ & 1.699 & 0.630 & 0.565 & 1.805 & 0.212 & 0.596 & 0.234 & 0.179 & 0.790 \\
1.0$\times10^{-4}$ & 1.699 & 0.630 & 0.565 & 1.805 & 0.212 & 0.596 & 0.234 & 0.179 & 0.790 \\
4.0$\times10^{-3}$ & 1.601 & 0.409 & 0.389 & 0.717 & 0.102 & 0.058 & 0.083 & 0.205 & 0.247 \\
8.0$\times10^{-3}$ & 1.699 & 0.365 & 0.365 & 0.500 & 0.095 & 0.076 & 0.076 & 0.153 & 0.184 \\
2.0$\times10^{-2}$ & 1.699 & 0.254 & 0.249 & 0.264 & 0.168 & 0.113 & 0.092 & 0.138 & 0.166 \\
\hline\hline
\end{tabular}
\end{center}
\end{table*}


\begin{landscape}
\begin{table}[!h]
{\small
\caption{Returned mass and Integrated Yields for $M_{up}=40$ \msun.}
\label{my40}
\begin{tabular}{cccccccccccc}
\hline
\hline
$Z_i$ & R & y$_H$ & y$_{He}$ & y$_C$ & y$_N$ & y$_O$ & y$_{Ne}$ & y$_S$ & y$_{Cl}$ &y$_{Ar}$ & y$_{Fe}$\\
\hline
LIMS&&&&&&&&&&&\\
\hline
1.0$\times10^{-5}$& 2.034$\times10^{-1}$ &-2.252$\times10^{-2}$&1.796$\times10^{-2}$&2.467$\times10^{-3}$&1.593$\times10^{-3}$&6.847$\times10^{-5}$&1.220$\times10^{-5}$&9.724$\times10^{-8}$& -     &-      & -9.154$\times10^{-8}$\\
1.0$\times10^{-4}$& 2.034$\times10^{-1}$&-2.252$\times10^{-2}$&1.796$\times10^{-2}$&2.467$\times10^{-3}$&1.593$\times10^{-3}$&6.847$\times10^{-5}$&1.220$\times10^{-5}$& 9.724$\times10^{-8}$& -     &-      &-9.154$\times10^{-8}$\\
4.0$\times10^{-3}$& 2.055$\times10^{-1}$&-1.175$\times10^{-2}$&9.680$\times10^{-3}$&1.357$\times10^{-3}$&5.579$\times10^{-4}$&-1.516$\times10^{-5}$&2.804$\times10^{-6}$&-2.678$\times10^{-7}$&-     &-      &-1.791$\times10^{-6}$\\
8.0$\times10^{-3}$& 2.099$\times10^{-1}$&-9.427$\times10^{-3}$&7.826$\times10^{-3}$&1.059$\times10^{-3}$&4.329$\times10^{-4}$&-3.745$\times10^{-5}$&1.568$\times10^{-6}$&-3.084$\times10^{-7}$&-     &-      &-2.109$\times10^{-6}$\\
2.0$\times10^{-2}$& 2.132$\times10^{-1}$&-5.768$\times10^{-3}$&5.115$\times10^{-3}$&2.073$\times10^{-4}$&4.059$\times10^{-4}$&-7.928$\times10^{-5}$&1.690$\times10^{-7}$&-1.538$\times10^{-7}$&-     &-      &-1.127$\times10^{-6}$\\
\hline
MS &&&&&&&&&&&\\
\hline
 1.0$\times10^{-8}$& 7.975$\times10^{-1}$  &-2.367$\times10^{-1}$& 1.412$\times10^{-1}$& 2.451$\times10^{-2}$& 1.787$\times10^{-3}$& 4.602$\times10^{-2}$& 7.030$\times10^{-3}$& 1.645$\times10^{-3}$& 4.226$\times10^{-6}$& 3.318$\times10^{-4}$& 7.735$\times10^{-3}$\\
 1.0$\times10^{-5}$& 5.287$\times10^{-2}$&-1.255$\times10^{-2}$& 8.408$\times10^{-3}$& 5.677$\times10^{-4}$& 8.739$\times10^{-6}$& 1.887$\times10^{-3}$& 4.606$\times10^{-4}$&  1.489$\times10^{-4}$& 5.211$\times10^{-7}$& 3.165$\times10^{-5}$& 5.539$\times10^{-4}$\\
 1.0$\times10^{-4}$& 5.289$\times10^{-2}$&-1.276$\times10^{-2}$& 8.310$\times10^{-3}$& 5.457$\times10^{-4}$& 3.400$\times10^{-5}$& 2.131$\times10^{-3}$& 4.370$\times10^{-4}$& 1.500$\times10^{-4}$& 8.336$\times10^{-7}$& 3.236$\times10^{-5}$& 6.114$\times10^{-4}$\\
 4.0$\times10^{-3}$& 5.289$\times10^{-2}$&-1.307$\times10^{-2}$& 8.127$\times10^{-3}$& 5.110$\times10^{-4}$& 7.448$\times10^{-5}$& 2.515$\times10^{-3}$& 3.997$\times10^{-4}$& 1.517$\times10^{-4}$& 1.336$\times10^{-6}$& 3.358$\times10^{-5}$& 7.042$\times10^{-4}$\\
 8.0$\times10^{-3}$& 5.287$\times10^{-2}$&-1.336$\times10^{-2}$& 7.655$\times10^{-3}$& 7.458$\times10^{-4}$& 1.309$\times10^{-4}$& 2.974$\times10^{-3}$& 4.483$\times10^{-4}$& 1.493$\times10^{-4}$& 1.250$\times10^{-6}$& 3.041$\times10^{-5}$& 5.971$\times10^{-4}$\\
 2.0$\times10^{-2}$& 5.282$\times10^{-2}$&-1.386$\times10^{-2}$& 6.922$\times10^{-3}$& 1.120$\times10^{-3}$& 2.211$\times10^{-4}$& 3.707$\times10^{-3}$& 5.261$\times10^{-4}$& 1.458$\times10^{-4}$& 1.109$\times10^{-6}$& 2.531$\times10^{-5}$& 4.258$\times10^{-4}$\\
\hline
SNIa$^a$ &&&&&&&&&&&\\
\hline
-&- & -                             &-                                                                              & 1.457$\times10^{-5}$& 3.501$\times10^{-10}$&  4.315$\times10^{-5}$ & 6.096$\times10^{-7}$&  2.538$\times10^{-5}$&   4.044$\times10^{-8}$& 3.802$\times10^{-12}$& 1.850$\times10^{-4}$ \\
\hline
\multicolumn{7}{l}{$^a$ SNIa yields are independent of the initial metallicity.}\\
\hline
\hline
\end{tabular}
}
\end{table}
\end{landscape}


\begin{landscape}
\begin{table}[!h]
{\small
\caption{Returned mass and Integrated Yields assuming $M_{up}=80$ \msun.}
\label{my80}
\begin{tabular}{cccccccccccc}
\hline
\hline
$Z_i$ & R & y$_H$ & y$_{He}$ & y$_C$ & y$_N$ & y$_O$ & y$_{Ne}$ & y$_S$ & y$_{Cl}$ &y$_{Ar}$ & y$_{Fe}$\\
\hline
LIMS&&&&&&&&&&&\\
\hline
1.0$\times10^{-5}$&2.018$\times10^{-1}$&-2.234$\times10^{-2}$&1.781$\times10^{-2}$&2.447$\times10^{-3}$&1.579$\times10^{-3}$&6.790$\times10^{-5}$&1.210$\times10^{-5}$& 9.641$\times10^{-8}$& -     &-      & -9.078$\times10^{-8}$\\
1.0$\times10^{-4}$&2.018$\times10^{-1}$&-2.234$\times10^{-2}$&1.781$\times10^{-2}$&2.447$\times10^{-3}$&1.579$\times10^{-3}$&6.790$\times10^{-5}$&1.210$\times10^{-5}$& 9.641$\times10^{-8}$& -     &-      & -9.078$\times10^{-8}$\\
4.0$\times10^{-3}$&2.039$\times10^{-1}$&-1.165$\times10^{-2}$&9.600$\times10^{-3}$&1.346$\times10^{-3}$&5.531$\times10^{-4}$&-1.503$\times10^{-5}$&2.780$\times10^{-6}$&-2.655$\times10^{-7}$&-     &-      &-1.776$\times10^{-6}$\\
8.0$\times10^{-3}$&2.082$\times10^{-1}$&-9.349$\times10^{-3}$&7.761$\times10^{-3}$&1.050$\times10^{-3}$&4.292$\times10^{-4}$&-3.713$\times10^{-5}$&1.555$\times10^{-6}$&-3.057$\times10^{-7}$&-     &-      &-2.091$\times10^{-6}$\\
2.0$\times10^{-2}$&2.115$\times10^{-1}$&-5.720$\times10^{-3}$&5.073$\times10^{-3}$&2.055$\times10^{-4}$&4.025$\times10^{-4}$&-7.861$\times10^{-5}$&1.676$\times10^{-7}$&-1.525$\times10^{-7}$&-     &-      &-1.117$\times10^{-6}$\\
\hline
MS &&&&&&&&&&&\\
\hline
 1.0$\times10^{-8}$& 8.686$\times10^{-1}$& -2.794$\times10^{-1}$& 1.535$\times10^{-1}$& 3.115$\times10^{-2}$& 2.351$\times10^{-3}$& 6.784$\times10^{-2}$& 8.776$\times10^{-3}$& 1.570$\times10^{-3}$& 4.052$\times10^{-6}$& 3.167$\times10^{-4}$& 7.244$\times10^{-3}$\\
 1.0$\times10^{-5}$& 6.089$\times10^{-2}$&-1.668$\times10^{-2}$& 9.997$\times10^{-3}$& 6.676$\times10^{-4}$& 9.123$\times10^{-6}$& 3.865$\times10^{-3}$& 8.106$\times10^{-4}$& 1.492$\times10^{-4}$& 5.208$\times10^{-7}$& 3.164$\times10^{-5}$& 5.628$\times10^{-4}$\\
 1.0$\times10^{-4}$& 6.095$\times10^{-2}$&-1.686$\times10^{-2}$& 9.759$\times10^{-3}$& 6.552$\times10^{-4}$& 3.807$\times10^{-5}$& 4.226$\times10^{-3}$& 7.751$\times10^{-4}$& 1.493$\times10^{-4}$& 8.416$\times10^{-7}$& 3.215$\times10^{-5}$& 6.192$\times10^{-4}$\\
 4.0$\times10^{-3}$& 6.103$\times10^{-2}$&-1.712$\times10^{-2}$& 9.350$\times10^{-3}$& 6.357$\times10^{-4}$& 8.446$\times10^{-5}$& 4.791$\times10^{-3}$& 7.188$\times10^{-4}$& 1.494$\times10^{-4}$& 1.357$\times10^{-6}$& 3.304$\times10^{-5}$& 7.105$\times10^{-4}$\\
 8.0$\times10^{-3}$& 6.110$\times10^{-2}$&-1.727$\times10^{-2}$& 9.075$\times10^{-3}$& 1.075$\times10^{-3}$& 1.539$\times10^{-4}$& 4.749$\times10^{-3}$& 7.163$\times10^{-4}$& 1.469$\times10^{-4}$& 1.286$\times10^{-6}$& 2.983$\times10^{-5}$& 5.987$\times10^{-4}$\\
 2.0$\times10^{-2}$& 6.119$\times10^{-2}$&1.754$\times10^{-2}$& 8.659$\times10^{-3}$& 1.776$\times10^{-3}$& 2.649$\times10^{-4}$& 4.689$\times10^{-3}$& 7.120$\times10^{-4}$& 1.431$\times10^{-4}$& 1.168$\times10^{-6}$& 2.467$\times10^{-5}$& 4.197$\times10^{-4}$\\
\hline
SNIa$^a$ &&&&&&&&&&&\\
\hline
-&- & - &-& 1.445$\times10^{-5}$& \tiny  3.470$\times10^{-10}$&   4.278$\times10^{-5}$& 6.043$\times10^{-7}$&  2.516$\times10^{-5}$&   4.009$\times10^{-8}$& \tiny  3.769$\times10^{-12}$& 1.834$\times10^{-4}$ \\
\hline
\multicolumn{7}{l}{$^a$ SNIa yields are independent of the initial metallicity.}\\
\hline
\hline
\end{tabular}
}
\end{table}
\end{landscape}



\begin{thebibliography}{} 

\bibitem[]{all98} Allen, C., Carigi, L., \& Peimbert, M. 1998, 
\apj, 494, 247 

\bibitem[]{buz06} Buzzoni, A., Arnaboldi, M., \& Corradi, R. L. M. 2006, 
\mnras, 368, 877 

\bibitem[]{car06} Carigi, L., Col{\'\i}n, P., \& Peimbert, M. 2006, 
\apj, 644, 924 (CCP06) 

\bibitem[]{car08} Carigi, L., \& Hern\'andez, X. 2008,
\mnras, 390, 582 

\bibitem[]{car11} Carigi, L.,\& Peimbert, M. 2011, 
RMxAA, 47, 139

\bibitem[]{car06} Carigi, L., Peimbert, M.,  Esteban, C., Garc{\'\i}a- Rojas, J.,  2005, 
\apj, 623, 213

\bibitem[]{chiap05} Chiappini, C., Romano, D \& Matteucci, F., 2003
\mnras, 339, 63.

\bibitem[]{deblok00} de Blok, W. J. G.,  \& Walter, F., 2000, 
\apj, 537, 95.

\bibitem[]{deblok03} de Blok, W. J. G., \& Walter, F., 2003, 
\mnras, 341, 39.

\bibitem[]{del09} Delgado-Inglada, G., Rodr{\'\i}guez, M., Mampaso, A., \& Viironen, K. 2009, 
\apj, 694, 1335 

\bibitem[]{dun09} Dunkley, J., Komatsu, E., Nolta, M. R., et al. 2009, \apjs, 180, 306.

\bibitem[]{est04} Esteban, C. Peimbert, M., Garc{\'\i}a- Rojas, J., 
Ruiz, M. T., Peimbert, A., \& Rodr{\'\i}guez, M. 2004, 
\mnras, 355, 229 

\bibitem[]{fra08} Franco, I., \&  Carigi, L. 2008, 
RMxAA, 44, 311 

\bibitem[gieren]{gie06} Gieren, W., Pietrzy\'nski, G., Nalewajko, K., et al. 2006, 
\apj, 647, 1056

\bibitem[]{goo05} Goodwin, S. P., \& Pagel, B. E. J. 2005, 
\mnras, 359, 707 

\bibitem[]{gou10} Gouliermis, D. A., Schmeja, S., Klessen, R. S., de Blok, W. J. G., \&  Fabian, W. 2010, 
\apj, 725, 1717

\bibitem[]{her09a} Hern\'andez-Mart{\'\i}nez, L. 2010, Ph.D. Thesis,
Universidad Nacional Aut\'onoma de M\'exico.  

\bibitem[]{her09} Hern\'andez-Mart{\'\i}nez, L., Pe\~na, M., Carigi, L., \& Garc{\'\i}a-Rojas, J. 2009, 
\aap, 505 1027 (HPCG09) 

\bibitem[]{hir07} Hirschi, R. 2007,
\aap, 461, 571 

\bibitem[]{hir05} Hirschi, R., Meynet, G., \&  Maeder, A. 2005, 
\aap, 433, 1013 

\bibitem[]{izo07} Izotov, Y. I., Thuan, T. X.,
\& Stasi\'nska, G. 2007, \apj, 662, 15 

\bibitem[]{kar07} Karakas, A., \& Lattanzio, J. C. 2007, 
PASA  24, 103 

\bibitem[]{kob11} Kobayashi, C., Karakas, A. I., Umeda, H. 2011
\mnras, 414, 3231

\bibitem[]{kob06}
Kobayashi, C., Umeda, H., Nomoto, K., Tominaga, N., Ohkubo, T. 2006
\apj 653,  1145

\bibitem[]{kro93} Kroupa, P., Tout, C. A., \& Gilmore, G. 1993, 
\mnras, 262, 545 

\bibitem[]{liu06} Liu, X.-W., Barlow, M. J., Zhang, Y., Bastin, R. J., \& Storey, P.J. 2006, 
\mnras, 368, 1959 

\bibitem[]{liu04} Liu, Y., Liu, X.-W., Barlow, M. J., \& Lou, S.-G. 2004, 
\mnras, 353, 1251 

\bibitem[]{mac06} Maciel, W. J., Lago, L. G. \& Costa, R. D. D., 2006, 
\aap, 453, 587

\bibitem[Maeder(1992)]{mae92} Maeder,  A. 1992, 
\aap, 264, 105

\bibitem[]{man08} Mannucci, F. 2008, 
ChJAS, 8, 143 

\bibitem[]{mey02} Meynet, G., \&  Maeder, A. 2002, 
\aap, 390, 561

\bibitem[]{mol06} Moll\'a, M., V\'ilchez, J. M., Gavil\'an, M.; D\'iaz, A. I. 2006
\mnras 372,  1069 

\bibitem[]{nom97} Nomoto, K., Iwamoto, K., Nakasato, N. et al. 1997, 
NuPhA, 621, 467

\bibitem[]{pag89} Pagel, B. E. J. 1989, 
RMxAA, 18, 161

\bibitem[]{pei10} Peimbert, A., \& Peimbert, M. 2010, 
\apj, 724, 791 

\bibitem[]{pei05} Peimbert, A., Peimbert, M., \&  Ruiz, M. T. 2005, 
\apj, 634, 1056

\bibitem[]{pei07a} Peimbert, M., Luridiana, V., \& Peimbert, A. 2007a, 
\apj, 666, 636 

\bibitem[]{pei07b}
Peimbert, M., Luridiana, V., Peimbert, A. \& Carigi, L. 2007b, 
ASPC, 374, 81   

\bibitem[]{pei95b} Peimbert, M., Luridiana, V. \& Torres-Peimbert, S. 1995b,
RMxAA, 31, 147

\bibitem[]{pei11} Peimbert, M.,  \& Peimbert, A. 2011, 
RMxAASC, in press (arXiv0912.3781).


\bibitem[]{pei95} Peimbert, M., Torres-Peimbert, S., \& Luridiana, V. 1995a, 
RMxAA, 31, 131 


\bibitem[]{pei92} Peimbert, M., Torres-Peimbert, S., Ruiz, M. T. 1992 
RMxAA, 24, 155

\bibitem[Portinari et al.(1998)]{por98} Portinari, L., Chiosi, C., \& Bressan, A. 1998, 
\aap, 334, 505

\bibitem[]{ric07} Richer, M. G., \&  McCall, M. 2007, 
\apj, 658, 328 

\bibitem[]{rol94} Rola, C., \& Stasi\'nska, G. 1994, 
\aap, 282, 199 

\bibitem[]{rom10} 	Romano, D., Karakas, A. I., Tosi, M., \& Matteucci, F. 2010
\aap, 522, 32

\bibitem[]{ser83} Serrano, A., \& Peimbert, M., 1983, 
RMxAA, 8, 117 

\bibitem[]{shi78} Shields, G. A. 1978,
\apj, 219, 559 

\bibitem[]{sim11} Sim\'on-D{\'\i}az, S., \& Stasi\'nska, G. 2011, 
\aap, 526, A48 

\bibitem[]{tal71} Talbot, R. J. Jr., \& Arnett, W. D. 1971, 
\apj, 170, 409 

\bibitem[]{tim95}
Timmes, F. X., Woosley, S. E., Weaver, Thomas A. 1995
ApJSS 98, 617

\bibitem[]{tin74} Tinsley, B. M. 1974, 
\apj, 192, 629 

\bibitem[]{tsa04} Tsamis, Y. G., Barlow, M. J., Liu, X.-W., Storey, P. J., \& Danziger, I. J. 2004, 
\mnras, 353, 953 

\bibitem[]{tsa11} Tsamis, Y. G., Walsh, J. R., V{\'\i}lchez, J. M., \& P\'equignot, D. 2011,
\mnras, 412, 1367 

\bibitem[]{ven01} Venn, K. A.,  Lennon, D. J., Kaufer, A., et al. 2001, 
\apj, 625, 754 

\bibitem[]{wes05} Wesson, R., Liu, X.-W., \& Barlow, M. J., 2005, 
\mnras, 362, 424

\bibitem[]{woo95} Woosley, S. E., \&   Weaver, T. A. 1995, 
\apjs, 101, 181

\end{thebibliography}
\end{document}